# Probing the ideal limit of interfacial thermal conductance in two-dimensional van der Waals heterostructures

Ting Liang,[1] Ke Xu,[1] Penghua Ying,[2] Wenwu Jiang,[3] Meng Han,[4] Xin Wu,[5] Wengen Ouyang,[3] Yimin Yao,[4] Xiaoliang Zeng,[4] Zhenqiang Ye,[6, *] Zheyong Fan,[7, †] and Jianbin Xu[1, ‡]

[1]*Department of Electronic Engineering and Materials Science and Technology Research Center, The Chinese University of Hong Kong, Shatin, N.T., Hong Kong SAR, 999077, P. R. China*
[2]*Department of Physical Chemistry, School of Chemistry, Tel Aviv University, Tel Aviv, 6997801, Israel*
[3]*Department of Engineering Mechanics, School of Civil Engineering, Wuhan University, Wuhan, Hubei 430072, China*
[4]*Shenzhen Institute of Advanced Electronic Materials, Shenzhen Institute of Advanced Technology, Chinese Academy of Sciences, Shenzhen 518055, PR China*
[5]*Institute of Industrial Science, The University of Tokyo, Tokyo 153-8505, Japan*
[6]*College of Materials Science and Engineering, Shenzhen University, Shenzhen 518055, China*
[7]*College of Physical Science and Technology, Bohai University, Jinzhou 121013, P. R. China*
(Dated: February 19, 2025)

Probing the ideal limit of interfacial thermal conductance (ITC) in two-dimensional (2D) heterointerfaces is of paramount importance for assessing heat dissipation in 2D-based nanoelectronics. Using graphene/hexagonal boron nitride (Gr/$h$-BN), a structurally isomorphous heterostructure with minimal mass contrast, as a prototype, we develop an accurate yet highly efficient machine-learned potential (MLP) model, which drives nonequilibrium molecular dynamics (NEMD) simulations on a realistically large system with over 300,000 atoms, enabling us to report the ideal limit range of ITC for 2D heterostructures at room temperature. We further unveil an intriguing stacking-sequence-dependent ITC hierarchy in the Gr/$h$-BN heterostructure, which can be connected to moiré patterns and is likely universal in van der Waals layered materials. The underlying atomic-level mechanisms can be succinctly summarized as energy-favorable stacking sequences facilitating out-of-plane phonon energy transmission. This work demonstrates that MLP-driven MD simulations can serve as a new paradigm for probing and understanding thermal transport mechanisms in 2D heterostructures and other layered materials.

## I. INTRODUCTION

Van der Waals heterostructures, formed by vertically stacking various monolayer two-dimensional (2D) materials, have emerged as a versatile platform for exploring fundamental physical phenomena and offering great opportunities for fabricating advanced electronic devices [1, 2]. Beyond their remarkable functional capabilities, the intrinsic thermal transport properties of 2D-based heterostructures have attracted considerable attention, as efficient heat dissipation is paramount to ensuring reliable device performance and long-term operational stability [3, 4]. Particularly, given that "the interface is the devil", yet it is omnipresent in devices, a pivotal and highly compelling question arises regarding the ideal upper limit of interfacial thermal conductance (ITC) in a pristine, atomically perfect 2D van der Waals heterostructure [5–12].

Among the various members of this family, the graphene/hexagonal boron nitride (Gr/$h$-BN) heterostructure stands out due to its vanishingly small lattice mismatch and the close structural and mass similarities between Gr and $h$-BN [13–15], making it a prominent candidate for achieving the highest ITC among 2D heterostructures. Therefore, the challenge can be reframed as exploring the ideal limit of ITC in Gr/$h$-BN heterostructure.

However, considerable discrepancies remain between theoretical predictions and experimental measurements of the ITC in Gr/$h$-BN heterostructure. The experimental values range from approximately 7 to 52 MW m$^{-2}$ K$^{-1}$ [6–8, 16], while theoretical predictions span a much broader range, from 5 to 442 MW m$^{-2}$ K$^{-1}$ [17–21]. Due to the limited resolution of thermal measurement techniques such as time-domain thermoreflectance (TDTR) and photothermal Raman [6–8, 16], experimentally isolating the ITC of 2D heterostructures is particularly challenging. In light of this, molecular dynamics (MD) simulations have emerged as the predominant theoretical approach for predicting the ITC values and providing atomic-scale mechanistic insights. This method naturally incorporates full anharmonicity and scales computationally linearly with system size, enabling large-scale and long-duration simulations, making it an ideal approach for considering the slight lattice mismatch between Gr and $h$-BN and the accompanying surface reconstruction-driven moiré patterns [13, 14, 22, 23].

The accuracy of MD simulations, however, depends heavily on the reliability of the interatomic potentials. For the Gr/$h$-BN heterostructure system, the commonly employed combination of Tersoff [24] and Lennard-Jones (LJ) [25] potentials have proven inadequate in accurately capturing intralayer and interlayer interactions [26, 27]. While the anisotropic interlayer potential (ILP) [27–31], parameterized using state-of-the-art density functional theory (DFT), can simultaneously capture the adhesion and anisotropic repulsive characteristics of the interlayer interaction, the intralayer interactions still rely on empirical potentials such as Tersoff [24] and adaptive intermolecular reactive empirical bond order (AIREBO) [32]. These limitations have prevented the ideal limit of ITC in the Gr/$h$-BN heterostructure from being fully explored, and the underlying moiré-pattern-related thermal transport mechanisms remain elusive–all of which are essential to provide calibrations for experimental measurements.

---

* yezq@szu.edu.cn
† brucenju@gmail.com
‡ jbxu@ee.cuhk.edu.hk

In this work, based on the neuroevolution potential (NEP) approach [33–35], we develop an accurate yet highly efficient machine-learned potential (MLP) specifically tailored for Gr/$h$-BN heterostructure systems and use it to reconstruct the experimentally observed moiré patterns and periods accurately. Employing NEP-driven nonequilibrium MD (NEMD) simulations on a large Gr/$h$-BN system comprising over 300,000 atoms, augmented with quantum-statistical corrections to the classical MD outcomes, the ITC values in quantitative agreement with experimental ones are achieved. Such an agreement, combined with the phonon density of state (PDOS) overlap criterion, enables us to report the ideal limit range of ITC for 2D heterostructures at room temperature. The quantum-statistical corrections derived from spectral thermal conductance also elucidate a moderate temperature dependence of ITC in the Gr/$h$-BN heterostructure.

Further, in the Gr/$h$-BN heterostructure, we identify a novel physical picture of stacking-sequence-dependent ITC (ABp > DW > AB > AA). To elucidate the atomic-level mechanisms, van der Waals potential energy, interlayer distances, and energy-weighted phonon dispersion are systematically calculated. Their results collectively reveal that stacking sequences with larger interlayer distances are unfavorable for out-of-plane phonon energy transmission, manifested by smaller group velocities and phonon lifetimes, thereby driving the stacking-sequence-dependent ITC. This phenomenon can further be associated with moiré patterns, likely leading to moiré-pattern-dependent ITC.

## II. RESULTS

**An accurate and efficient NEP model for Gr/$h$-BN Heterostructure.** MLP represents an emerging and powerful technology that leverages the robust fitting capabilities of machine learning to accurately describe potential energy surfaces (PESs) across the entire chemical and structural space provided by first-principles calculations, such as DFT [36]. A well-trained MLP can predict physical quantities and drive atomistic simulations with computational efficiency approaching that of an empirical potential, while maintaining the high fidelity of DFT and far beyond its spatial and temporal scales. Due to its high computational efficiency, the NEP framework has become a popular MLP method widely employed in thermal transport studies [37–42], and is therefore utilized in this work.

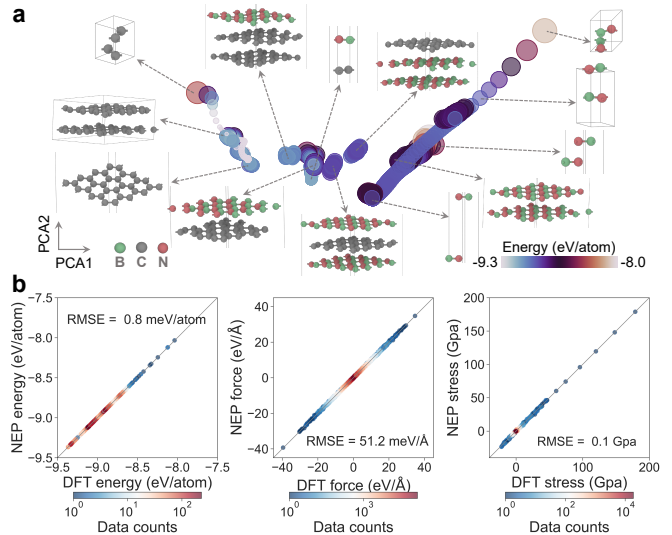

FIG. 1. **Training dataset and results for constructing the NEP model. a** A sketch map consisting of 4099 structures from the Gr/$h$-BN training dataset. Each colored dot represents an individual atomic configuration, with color indicating the energy associated with each structure. Snapshots of key configurations are visualized to highlight the diversity of the dataset. **b** Comparison between the NEP predictions and DFT reference data of energy, force, and stress for the training dataset. In the parity plot, the color intensity visualizes the distribution and density of the training data.

Constructing a high-quality and diverse training dataset is a pivotal step in the training NEP model. As shown in the sketch-map representation in Fig. 1a, the dataset encompasses various configurations essential for building the NEP model, such as Gr and $h$-BN monomers with different compositions, and the Gr/$h$-BN heterostructure with varying numbers of layers. Each point in the sketch map represents a distinct structure, with similar structures clearly grouped and dissimilar ones well separated. The spatial positions are determined through principal component analysis (PCA) of learned descriptors of the local atomic environments [43, 44], highlighting the breadth of the configurational space covered by the dataset. Detailed information on the creation and composition of the dataset can be found in the Methods section.

The initial structures were subjected to single-point DFT calculations to label the total energy, force, and stress. We employed the Perdew–Burke–Ernzerhof (PBE) functional [47] to describe the exchange-correlation interactions in the structure, and utilized the advanced many-body dispersion (MBD) framework [48, 49] to capture long-range dispersion interactions.

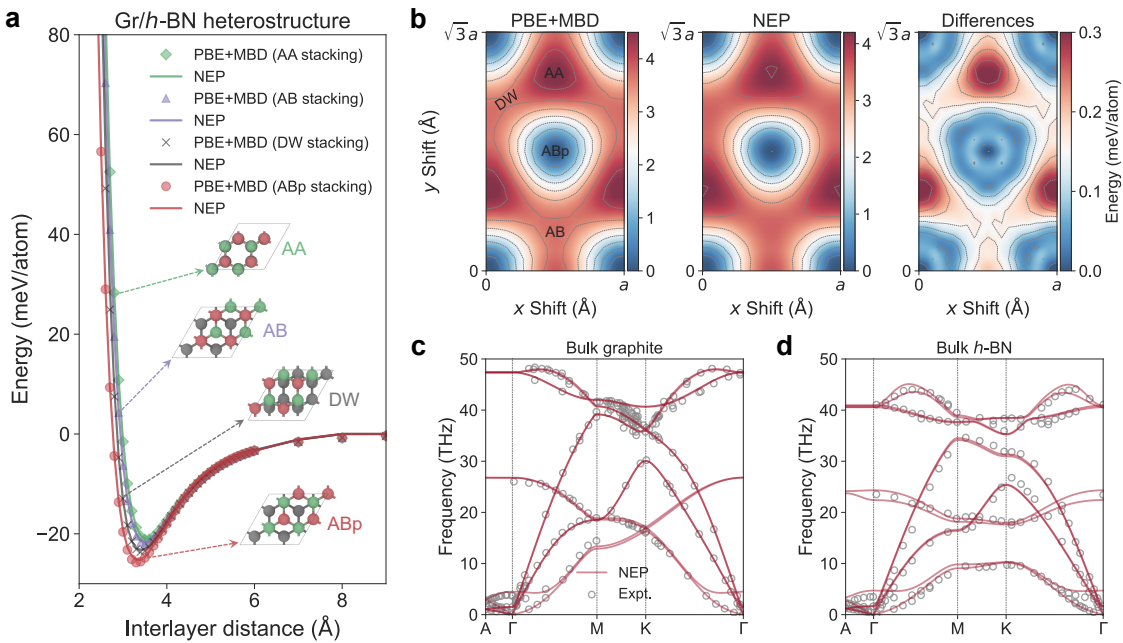

FIG. 2. **Various benchmarks for validating the NEP model.** **a** Comparison of binding energy curves for different stacked Gr/$h$-BN systems, calculated using DFT and NEP model. Atomic snapshots corresponding to different stacking types are visualized. **b** Comparison of sliding PES predicted by DFT and NEP model for the Gr/$h$-BN heterostructure. The AA, AB, DW, and ABp stacking modes are marked at the corresponding positions on the PES map. **c–d** Comparison of the phonon dispersion between experimental measurements and NEP model predictions, with experimental data for bulk graphite and $h$-BN extracted from Refs. [45] and [46], respectively.

Thus, the DFT reference data are labeled as PBE+MBD (for details, see Methods). Fig. 1b shows a well-trained NEP model for the Gr/$h$-BN heterostructure (see Methods for details on training and hyperparameters), which achieved relatively high accuracy. The root mean square errors (RMSEs) for energy, forces, and stress relative to the DFT references, are 0.8 meV/atom, 51.2 meV/Å, and 0.1 GPa, respectively.

To comprehensively validate the performance of the NEP model, we conducted a series of benchmark tests. As depicted in Fig. 2a, the NEP model accurately captures the PBE+MBD binding energy curves for different stacking types in the Gr/$h$-BN heterostructure, with ABp emerging as the most energetically favorable stacking configuration. Notably, the NEP model demonstrates a reliable prediction of domain wall (DW) stacking configurations with varying layer spacings without including them in the training set, highlighting its outstanding interpolation capability. The NEP model also accurately predicts the binding energy curves for different stacking types in both bulk and bilayer Gr and $h$-BN systems (see Figure S1), owing to its nearly error-free fitting of DFT energies (see Fig. 1b). For the shallow sliding PES of the Gr/$h$-BN heterostructure, the difference compared to PBE+MBD is less than 0.3 meV/atom (Fig. 2b). Furthermore, the phonon dispersion of bulk graphite and $h$-BN (Fig. 2c–d) calculated using the NEP model align well with experimental measurements [45, 46], despite minor deviations along certain high-symmetry paths. A detailed comparison of the sliding PESs and phonon dispersions for other systems against DFT references, provided in Figures S2 and S3, reveals minimal discrepancies, underscoring the ability of the NEP model to capture subtle differences in energy and force.

Utilizing the equilibrium lattice constants obtained from the NEP model, which are very close to those given by DFT calculations and experiments (see Supplemental Table S1 for the equilibrium lattice constants and interlayer spacing), we constructed a bilayer Gr/$h$-BN heterostructure unit cell with minimal lattice mismatch and in-plane dimensions of 13.8 × 23.9 nm (see Methods for details) to generate moiré patterns [13, 14, 22, 50]. The inherent lattice mismatch in the heterostructure causes interlayer atoms to form stacking sequences that are less energetically favorable. To accommodate the lattice disparity, during geometric optimization, atoms in each layer shift to minimize the system's energy, resulting in reconstructed surface regions where flat, near-optimal stacking areas are separated by narrow, elevated DWs [37, 51].

Fig. 3a–b illustrate the unit-cell moiré pattern and its corresponding supercell of the Gr/$h$-BN heterostructure optimized using the NEP model. Different regions display various stacking configurations that align well with the moiré patterns observed experimentally with scanning tunneling microscope (STM). The atomic reconstruction of the heterostructure leads to a loss of planarity, forming a strongly corrugated surface (Fig. 3c).

The moiré pattern periodicity from the NEP model (about 13.7 nm, see Fig. 3d) is well consistent with the results of the theoretical model at a zero-twist angle [13] and atomic force microscopy (AFM) measurements [22], further validating that the NEP model accurately captures the sliding energy landscape corrugation in the Gr/$h$-BN heterostructure (see Fig. 2a–b). Additionally, the empirical potentials Tersoff+ILP and Tersoff+LJ can reproduce the experimentally observed moiré patterns and their periodicities (refer to Figure S4). The LJ term (LJ parameters can be found in Ref. [52]), however, significantly underestimates the overall sliding energy landscape corrugation in Gr and $h$-BN [26], leading to substantial suppression of the height fluctuations in the moiré pattern obtained with Tersoff+LJ potential (see Figure S4b).

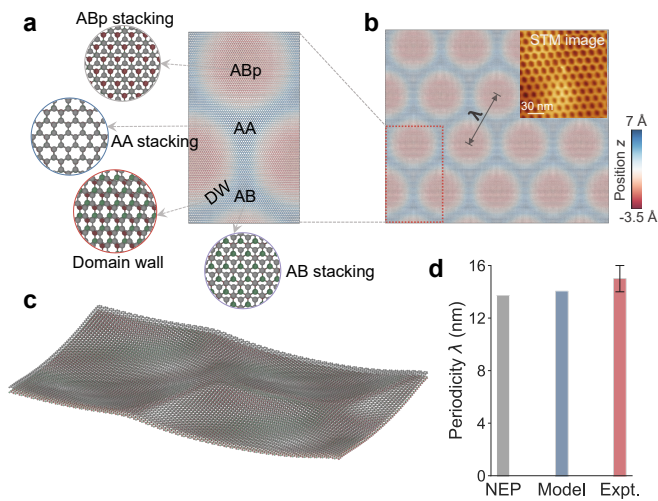

FIG. 3. **Moiré patterns and periodicities of the Gr/$h$-BN heterostructure.** **a** The unit-cell moiré pattern obtained from geometric optimization using the NEP model, containing AA, AB, DW, and ABp stacking types. **b** The supercell of the moiré patterns. The color bar represents the $z$-coordinate of the atoms, showing the undulating ripples. The upper right corner shows the moiré patterns observed experimentally using the STM [14]. **c** Schematic diagram of the free-standing bilayer Gr/$h$-BN heterostructure. **d** Comparison of the periodicity (wavelength, $\lambda$) of moiré patterns obtained through different methods. The modeling result [13] corresponds to the Gr/$h$-BN heterostructure with a zero-degree twist angle, while the experimental data are derived from AFM measurements [22].

Further, the NEP model is employed to perform geometric optimization of a bilayer $h$-BN with a 0.385° twist angle [53], successfully replicating the shape of the experimentally observed moiré pattern [50] and highlighting the stacking types in different regions (see Figure S5). This achievement demonstrates the strong generalization capability of the NEP model, even without prior training on twisted bilayer $h$-BN configurations.

For MD simulations, our NEP model, capable of capturing long-range dispersion interactions, achieves an impressive computational speed of $9.4 \times 10^6$ atom-step per second on four Nvidia GeForce RTX 4090 GPU cards (see Figure S6), closely rivaling the performance of the Tersoff+LJ empirical potential, which attains $1.3 \times 10^7$ atom-step per second using 72 Intel Xeon Gold 6420 CPU cores. In contrast, the Tersoff+ILP potential, running on 108 Intel Xeon Gold 6420 CPU cores, achieves only $1.7 \times 10^5$ atom-step per second, as ILP requires frequent updates of neighboring atoms and employs a large radial cutoff. The fast and accurate NEP model provides foundations for a comprehensive study of thermal transport in Gr/$h$-BN heterostructure.

**Ideal Limit of Interfacial Thermal Conductance in 2D Heterostructures.** By employing the NEMD method (refer to the Methods section for details) on a realistically large system containing over 300,000 atoms, we calculated the ITC of the Gr/$h$-BN heterostructure, as presented in Table I and Fig. 4a. At around 300 K, the ITC values calculated from the Tersoff+LJ and Tersoff+ILP potentials deviate significantly and moderately, respectively, from photothermal Raman and TDTR experimental measurements, whereas the NEP model yields results that closely match the experimental ones. The distribution of ITC data for different potentials from 20 independent calculations is shown in Figure S9. Classical MD simulations, however, cannot inherently account for quantum Bose-Einstein statistics [39–41]. To address this limitation, we performed a proper quantum-statistical correction to the classical ITC based on spectral thermal conductance (see Methods), achieving a closer match with experimental values at 300 K (see Table I).

We also used the approach-to-equilibrium MD (AEMD) technique (Methods) to calculate the ITC of Gr/$h$-BN heterostructure, revealing a significant underestimation compared to the experimental ones (see Supplemental Note S4 and Figure S8), likely because some low-frequency phonons with long lifetimes do not contribute to thermal conductance in AEMD [54]. Additionally, we confirmed that the Gr/$h$-BN heterostructure system exhibits almost no thermal rectification (Figure S10).

TABLE I. Comparison of the ITC of the Gr/$h$-BN heterostructure at 300 K, as calculated using NEMD method driven by various potentials, with experimental results.

| Method | Potentials/Techniques | ITC (MW m$^{-2}$ K$^{-1}$) |
|---|---|---|
| NEMD | Tersoff+LJ [20] | 442 |
|  | Tersoff+LJ [21] | $270.0 \pm 0.7$ |
|  | Tersoff+LJ (This work) | $250.7 \pm 18.5$ |
|  | Tersoff+ILP (This work) | $98.0 \pm 7.6$ |
|  | NEP (Classical) | $62.5 \pm 10.8$ |
|  | NEP (Quantum) | $58.4 \pm 11.0$ |
| Expt. | Raman [6] | $52.2 \pm 2.1$ |
|  | TDTR [8] | $34.5^{+11.6}_{-7.4}$ |

Among the various 2D heterostructures formed by Gr,

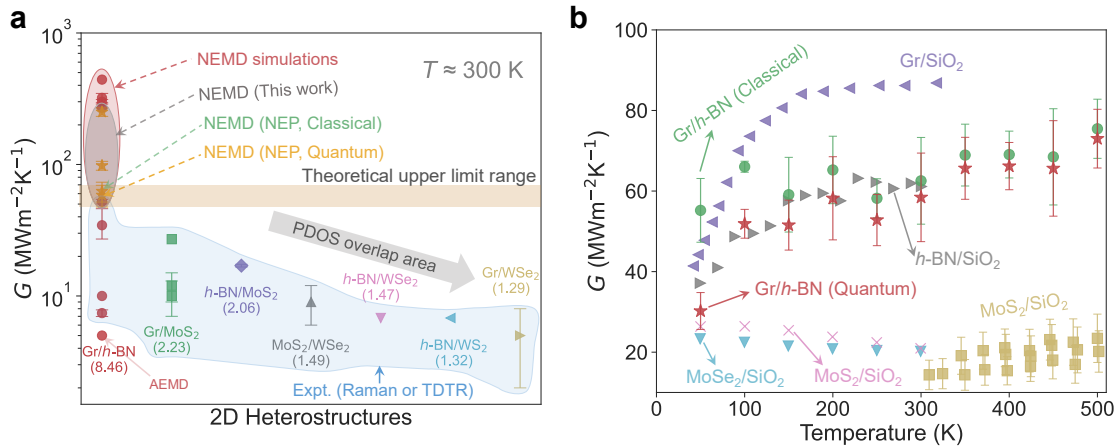

FIG. 4. **ITC data for various heterostructure systems. a** The correlation between ITC and PDOS overlap at room temperature for different 2D heterostructures. The light orange band emphasizes the ideal limit range of ITC for 2D heterostructures. The values of PDOS overlap (see Figure S12 for details) are marked below the labels of different systems, decreasing from left to right. The ITC data measured by Raman and TDTR experiments are highlighted with light blue shading, including Gr/$h$-BN [6–8, 16], Gr/$MoS_2$ [9, 10], $h$-BN/$MoS_2$ [6], $MoS_2$/$WSe_2$ [9], $h$-BN/$WSe_2$ [12], $h$-BN/$WS_2$ [12], and Gr/$WSe_2$ [9] heterostructures. The ITC for the Gr/$h$-BN heterostructure obtained from NEMD [20, 21], and AEMD [18] simulations are further compared with experimental ones. For clarity, the essential data are presented in Table I. **b** Temperature-dependent ITC for different heterostructure systems. For the Gr/$h$-BN heterostructure, the ITC includes the classical and quantum-corrected results from NEP calculations (with no available experimental data). For 2D/$SiO_2$ heterostructures, including Gr/$SiO_2$ [55], $h$-BN/$SiO_2$ [56], $MoS_2$/$SiO_2$ [57, 58], and $MoSe_2$/$SiO_2$ [58], the ITC values are based on experimental measurements.

$h$-BN, $MoS_2$, $WS_2$, and $WSe_2$, the Gr/$h$-BN heterointerface achieves the highest ITC (Fig. 4a), because Gr and $h$-BN have the highest similarity in structure and mass [15]. Conversely, the Gr/$WSe_2$ interface exhibits the lowest ITC, reflecting the largest disparity between the constituent materials, characterized by a substantial unit-cell mass mismatch (as shown in Figure S11) and differences in thermal properties. The relative "similarity" between a pair of 2D materials can be simplified to two key factors: the closeness of their average atomic masses (Figure S11), determining energy transmission probability, and the degree of overlap in their PDOSs (Figure S12), indicating the phonon modes available for energy transmission [5, 59]. Fig. 4a highlights the correlation between PDOS overlap and ITC across different 2D heterostructures, revealing a clear trend that greater overlap corresponds to higher ITC. This implies that the PDOS overlap can serve as a straightforward positive predictor for determining ITC.

In general, the intrinsic ITC of 2D heterostructures through van der Waals interactions is expected to have an upper limit, which we refer to here as the ideal limit of ITC. Approaching this ideal limit experimentally is fraught with challenges. The fabrication of a flawless interface is inherently difficult due to the susceptibility of 2D materials to impurity contamination and wrinkling during exfoliation and transfer. Moreover, the strong dependence of experimental outcomes on the "cleanliness" of the interface, coupled with the low measurement sensitivity of ITC [6–8, 16], further complicates accurate probing this upper bound. Using the Gr/$h$-BN system, which demonstrates the highest ITC among 2D heterostructures (see Fig. 4a and Supplemental Table S2), as a prototype, and employing NEP-driven NEMD simulations augmented with the PDOS overlap criterion, we determined the ideal limit of ITC for 2D heterostructures at room temperature to be $58.4 \pm 11.0$ MW m$^{-2}$ K$^{-1}$. In Fig. 4a, we highlight this ideal limit range, anticipating it to serve as a reference standard.

Fig. 4b compares the temperature-dependent ITC values for the Gr/$h$-BN heterostructure, as predicted by the NEP model, with experimental data for various 2D/$SiO_2$ heterostructures. Fundamentally, the temperature dependence of ITC arises from the temperature dependence of the heat capacity. Below the Debye temperature ($\Theta_D$), the population of phonon modes available for heat transfer across the interface increases with temperature. Similar to heat capacity, however, the ITC saturates and remains constant for $T > \Theta_D$. The "softer" material (with weaker bond strength and larger atomic mass) at the interface will ultimately limit the ITC. At temperatures above $\Theta_{D,\text{soft}}$, the high-energy phonon modes of the "harder" material remain populated but fail to transmit efficiently across the interface due to the absence of matching phonon modes in the "softer" material. This mechanism likely accounts for the weak temperature dependence of ITC in $MoS_2$/$SiO_2$ and $MoSe_2$/$SiO_2$, as the Debye temperatures of transition metal dichalcogenides are significantly lower than those of graphite and $h$-BN (see Figure S13).

As mentioned earlier, classical MD cannot account for quantum statistics, resulting in the ITC of the Gr/$h$-

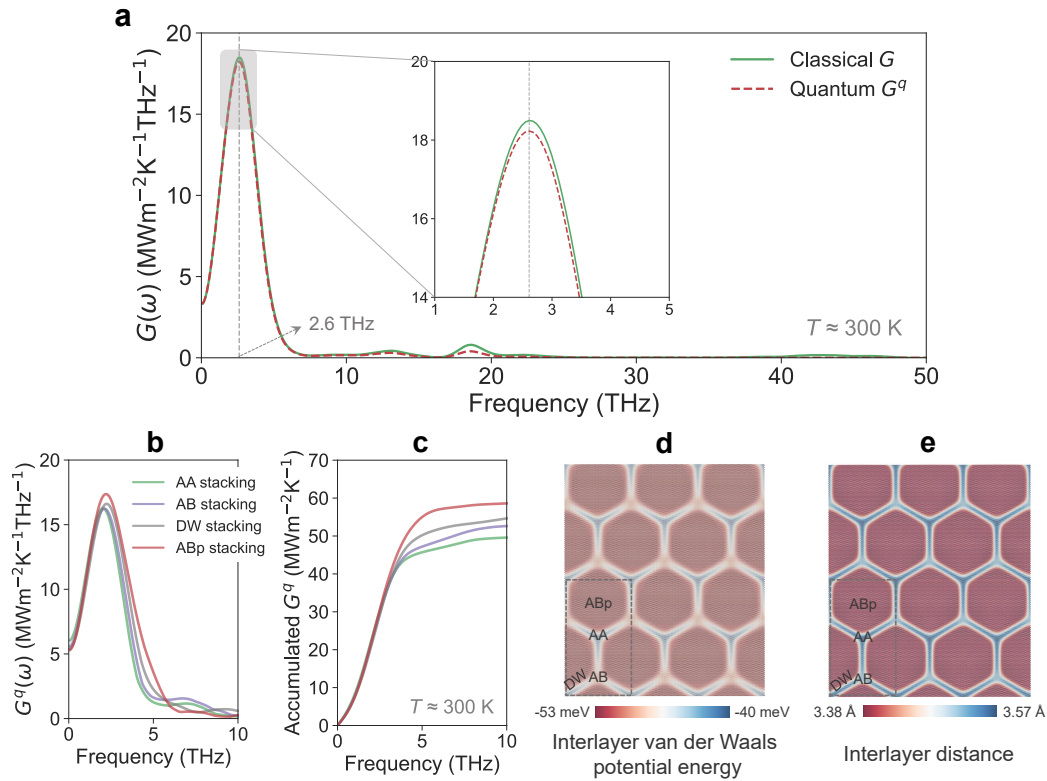

FIG. 5. **Spectral thermal conductance, interlayer energy and distance of Gr/$h$-BN heterostructure. a** Classical and quantum-corrected spectral thermal conductance of Gr/$h$-BN heterostructure at 300 K calculated using NEP. The inset shows a magnified view covering the 1–5 THz to provide a clearer comparison between the classical spectral thermal conductance and the quantum-corrected one. The gray dashed line emphasizes the central frequency of the peak for the spectral thermal conductance. **b–c** Quantum-corrected spectral thermal conductance and the corresponding accumulated thermal conductance for different stacking types. **d–e** Interlayer van der Waals potential energy (characterizing long-range dispersion interactions) and interlayer distance of the Gr/$h$-BN heterostructure. The distinct stacking types corresponding to different regions are highlighted, with the gray dashed lines representing the unit cell.

BN heterostructure calculated using the NEP model (depicted as circles in Fig. 4b) exhibiting an almost temperature-independent behavior. By incorporating quantum-statistical corrections based on spectral thermal conductance (see Methods and Figure S14), we reveal a moderate temperature dependence of the ITC (denoted by pentagrams), comparable to that of the $h$-BN/SiO$_2$ system. The temperature dependence of ITC for a broader range of heterostructures is comprehensively compared in Figure S15.

**Stacking-Sequence-Dependent Interfacial Thermal Transport.** It is now timely to delve into the quantum-statistical corrections based on spectral thermal conductance. As shown in Fig. 5a, the quantum-corrected spectral thermal conductance is slightly smaller than the classical one at 300 K, with substantial quantum-correction effects becoming apparent only at low temperatures (50 K, see Figure S14). This aligns with the moderate temperature dependence of the ITC demonstrated in Fig. 4b. The ITC of the Gr/$h$-BN heterostructure is predominantly driven by contributions from long-wavelength, low-frequency ($\omega/2\pi < 10$ THz) phonons. This finding is further corroborated by the spectral thermal conductance calculated using the empirical potentials Tersoff+ILP and Tersoff+LJ (Figure S16), which consistently highlight the dominant role of low-frequency phonons.

Furthermore, we evaluate the quantum-corrected spectral thermal conductance and their corresponding cumulative conductance for various stacking sequences at the Gr/$h$-BN heterointerface (see Methods), revealing a pronounced dependence of ITC on stacking sequences, as presented in Fig. 5b–c. The interlayer coupling strength and distance can act as straightforward metrics to interpret this phenomenon. The van der Waals potential energy and interlayer distance [31] obtained after geometric optimization (Fig. 5d–e) reveal that stacking sequences with lower/higher ground-state energies correspond to smaller/larger interlayer distances, thereby allowing more/less out-of-plane phonon energy to transmit through the heterointerface. For different stacking sequences, the interlayer potential energy and distance are ranked as ABp > DW > AB > AA, which corresponds to the ITC ranking of ABp > DW > AB > AA.

To further elucidate the stacking-sequence-dependent ITC behavior, it is instructive to examine the vibrational spectra along the Γ–A path [37, 60]. Using the spectral energy density (SED) technique, we calculated the energy-weighted phonon dispersion and derived the phonon lifetimes for various stacking configurations (see Methods). In Fig. 6a–c, a pronounced collapse (softening) of the transverse acoustic (TA) and longitudinal acoustic (LA) phonon modes, which dominate interfacial thermal transport [37, 60], is observed in the AA and AB stacking types compared to the ABp ones. Additionally, the ABp stacking type demonstrates larger group velocities (Figure S17) and longer phonon lifetimes (Fig. 6d) than the other two. Figure S18 presents the phonon lifetimes derived from the Lorentzian fitting of the SED, vividly showcasing the softening (redshift) of the TA and LA phonon modes when transitioning from ABp to AA stacking. These observations collectively indicate that energy-favorable stacking facilitates out-of-plane phonon energy transmission through reducing interlayer distances, manifesting in higher group velocities and prolonged phonon lifetimes, thereby driving the stacking-sequence-dependent ITC.

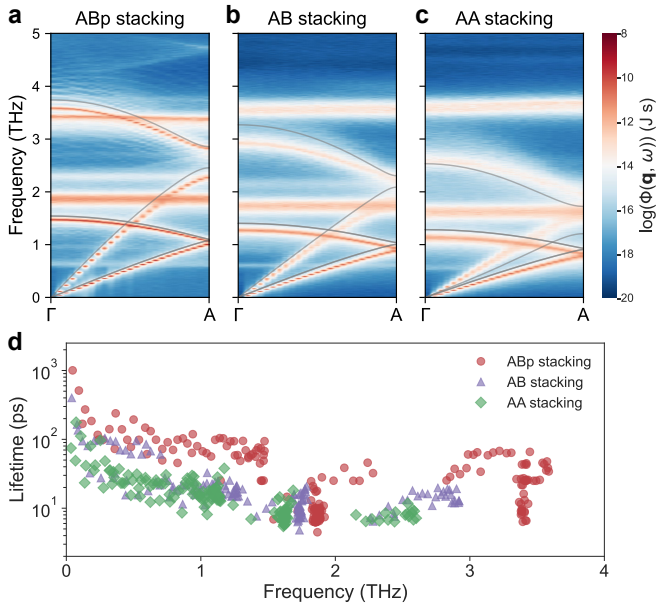

FIG. 6. **Phonon SED and the corresponding lifetimes.** **a–c** Phonon SED (from Γ to A, $\omega/2\pi < 5$ THz) for different stacking types of the Gr/$h$-BN heterostructure. The gray solid line represents lattice dynamics calculations at 0 K. **d** Phonon lifetimes for different stacking configurations, obtained through Lorentzian fitting of the SED peaks (see Figure S18).

We averaged the spectral thermal conductance calculated over the entire heterointerface and across all distinct stacking configurations, achieving quantitative agreement (Figure S19). In light of this, to obtain ITC values for 2D heterostructures from the NEMD+NEP method that are comparable to experimental measurements, it is insufficient to consider only a single stacking type in the computational model. Instead, at least one moiré period (see Fig. 3) containing multiple stacking types should be included. During the NEMD process at 300 K, the moiré pattern at the heterointerface remains clearly visible (see Figure S20).

The stacking-sequence-dependent ITC can be further correlated with moiré patterns. In 2D layered materials, the essence of moiré patterns stems from lattice mismatch caused by heterogeneous stacking, twisted angles, interlayer sliding, and heterostrain engineering. To minimize system energy, atomic reconstruction redistributes interlayer stacking sequences and induces diverse moiré patterns. It can be anticipated that different moiré patterns exhibit varying stacking proportions, leading to moiré-pattern-dependent ITC governed by distinct stacking sequences. This scenario invites a more comprehensive exploration of diverse layered materials.

### III. SUMMARY AND DISCUSSIONS

In summary, given the minimal lattice mismatch between Gr and $h$-BN, we use the Gr/$h$-BN system as a prototype to probe the ideal limit of ITC in 2D heterostructures. Building on the PBE functional combined with a state-of-the-art MBD correction, we develop a highly efficient NEP model for the Gr/$h$-BN heterostructure with near-first-principles accuracy. With this MLP model, we successfully reproduce the experimentally observed moiré patterns and periods, and perform NEMD simulations on a large Gr/$h$-BN heterostructure system with over 300,000 atoms. Through a proper quantum-statistical correction to the classical MD results, we achieve ITC values quantitatively consistent with experimental ones, thereby determining the ideal limit of ITC to be 58.4 ± 11.0 MW m$^{-2}$ K$^{-1}$ for 2D heterostructures at room temperature. This ideal limit is expected to serve as a standard reference for thermal design in 2D-based nanoelectronics.

Beyond this breakthrough, we unveil a novel stacking-sequence-dependent ITC hierarchy: ABp > DW > AB > AA in the Gr/$h$-BN heterostructure. The mechanism behind this picture can be attributed to stacking sequences with larger interlayer distances being unfavorable for out-of-plane phonon energy transmission, as evidenced by lower group velocities (the collapse of TA and LA phonon modes from ABp to AA stacking) and reduced phonon lifetimes, which ultimately drive the stacking-sequence-dependent ITC. This phenomenon can be further associated with moiré patterns, potentially giving rise to moiré-pattern-dependent ITC. Inspired by these observations, we consider that this stacking-sequence-dependent thermal transport scenario is likely universal in van der Waals layered materials and may extend to other properties such as ferroelectricity, spintronics, and magnetism [61–63].

## IV. METHODS

**Construction of Training Dataset.** Constructing the training structures of layered materials like Gr and $h$-BN for the NEP model is relatively straightforward, with sampling primarily from structural perturbations to capture diverse stress conditions and MD simulations to account for various temperature regimes [38]. Starting from the optimized structure, we introduced random cell deformations (-3% to 3%) and atomic displacements (within 0.1 Å) for structural perturbations. For MD simulation, DFT-driven MD simulation (see Supplemental Note S1 for details) is the main approach, and empirical potential-driven MD simulation is the auxiliary one.

The training dataset comprises four types of structures: monolayers, bilayers, trilayers, and bulk materials (see Fig. 1a for typical structures). For monolayers, we generated 50 perturbed structures each for Gr and $h$-BN and uniformly sampled 200 additional structures from each DFT-MD simulation, yielding a total of 500 monolayer structures. Using the same methods, we generated 500 bulk structures for graphite and $h$-BN. To accurately describe the binding energy curves, we varied the interlayer spacing to produce 59 additional bulk structures for graphite and $h$-BN, resulting in a total of 618 bulk configurations.

Due to the large number of atoms in trilayer structures (see Fig. 1a), we performed empirical potential-driven (Tersoff+ILP) [24, 28–31] MD simulations in the isothermal-isobaric (NpT) ensemble, generating 787 structures by linearly increasing the target temperature from 10 K to 1500 K over 200 ps.

For bilayer graphene, we considered AA and AB stackings (see Figure S1c) and generated 200 structures through perturbations and varying interlayer distances. In the case of bilayer $h$-BN, we explored five stacking types—AA, AAp, AB, ABp, and ApB (see Figure S1d)—producing 500 structures. Regarding Gr/$h$-BN heterostructure, we considered AA, AB, and ABp stackings (see Fig. 2a), producing a total of 900 structures using perturbations, DFT-driven MD, and varied interlayer spacings.

Additionally, rigid sliding configurations of bilayer Gr (AB stacking), bilayer $h$-BN (AA stacking), and bilayer Gr/$h$-BN heterostructure (AA stacking) were incorporated into the training set. Starting from the optimized configurations, we constructed a set of laterally shifted structures by rigidly translating the top layer [64]. In each case, a 6 × 11 grid of shifted configurations was generated along the zigzag and armchair directions, with each frame including two perturbed structures, yielding 198 structures per case and 594 configurations overall.

In total, we generated a comprehensive dataset of 4,099 structures (containing 229,760 atoms), all of which are utilized as the training set for the NEP model.

**DFT Calculations for Reference Data Generation.** After preparing the initial training configurations, the Vienna Ab initio Simulation Package (VASP) that implemented the projected augmented wave (PAW) method [65] and a plane-wave basis set [66] was used to perform quantum-mechanical calculations to obtain reference data, including the energy and virial of each structure, as well as the forces on each atom in the structure. The electronic exchange-correlation interaction was described by the PBE functional [47] within the generalized gradient approximation. Single-point calculations were performed with an energy cutoff of 850 eV and a Γ-centered $k$-point mesh, using a spacing of 0.15/Å to sample the Brillouin zone. The electronic self-consistent loop was converged to a threshold of $10^{-8}$ eV. For accurate total energy calculations, the tetrahedron method with Blöchl corrections was employed. To prevent periodic interactions in the $z$-direction, a vacuum layer of 50 Å thickness was applied. The state-of-the-art MBD correction [48, 49] was employed to accurately capture the long-range dispersion interactions. Details of the INCAR file used for generating the reference data are provided in Supplemental Note S2.

**The NEP Model Training.** We trained the NEP model for Gr/$h$-BN heterostructure by employing the fourth-generation (NEP4) scheme [35]. The NEP approach [33], based on a feedforward neural network (NN), uses a single hidden layer with $N_{\text{neu}}$ neurons to represent the site energy $U_i$ of atom $i$ as a function of a descriptor vector with $N_{\text{des}}$ components,

$$U_i = \sum_{\mu=1}^{N_{\text{neu}}} w_\mu^{(1)} \tanh\left(\sum_{\nu=1}^{N_{\text{des}}} w_{\mu\nu}^{(0)} q_\nu^i - b_\mu^{(0)}\right) - b^{(1)}, \quad (1)$$

where $\tanh(x)$ is the activation function, $\mathbf{w}^{(0)}$, $\mathbf{w}^{(1)}$, $\mathbf{b}^{(0)}$, and $b^{(1)}$ are the trainable weight and bias parameters in the NN. The local atom-environment descriptor $q_\nu^i$ is constructed from a set of radial and angular components [34].

Optimizing the NEP model parameters involves minimizing a loss function that combines a weighted sum of RMSEs for energy, force, and virial, along with regularization terms. After extensive benchmarking, we identified the optimal hyperparameters for NEP model training, available in the `nep.in` file (see Supplemental Note S3). To more accurately describe the binding energy curves and sliding PESs, we increased the weight of the 816 configurations for the four atoms in the training set from 1 to 8, highlighting their relative importance in the total loss function.

**Calculations of Interfacial Thermal Conductance.**
*NEMD simulations.* We primarily employed the NEMD method to calculate the ITC of the Gr/$h$-BN heterostructure. Given the inherent in-plane lattice mismatch of approximately 1.8% between Gr and $h$-BN [13, 14, 22, 23], we constructed supercells by expanding the Gr unit cell 56-fold and the $h$-BN unit cell 55-fold along both the armchair and zigzag directions. This resulted in in-plane dimensions of 13.8 × 23.9 nm, significantly reducing the in-plane size effects on ITC. Subsequently, we constructed a multilayer Gr/$h$-BN heterostructure model featuring a single heterointerface, comprising 12 layers of graphene (12,544 atoms per layer) and 13 layers of $h$-BN (12,100 atoms per layer), totaling 307,828 atoms. Through testing, this layer configuration effectively minimizes the impact of out-of-plane dimensions on ITC while maintaining computational efficiency. The initial interlayer spacing was set to 0.35 nm, and subsequently optimized during equilibration to a more physically realistic value. Detailed information on the model can be found in Figure S7.

The initial structure was first equilibrated for 0.5 ns under an NpT ensemble (with zero target pressure) at the target temperatures. This was followed by a 0.25 ns run in the NVT ensemble. We adopted a time step of 0.5 fs, which has been proven sufficiently small for accurate results. During the production stage lasting 2.5 ns, two Langevin local thermostats [67], specifically a heat source and sink (with a temperature difference of 60 K), are employed to generate a nonequilibrium steady state with constant heat flux. At this stage, a noticeable temperature jump $\Delta T$, occurs at the Gr/$h$-BN heterointerface (see Figure S7c), representing the average interfacial temperature difference between Gr and the neighboring $h$-BN layer. One can calculate the ITC ($G$) according to Fourier's law:

$$G = \frac{dE/dt}{S\Delta T}, \quad (2)$$

where $S$ is the cross-sectional area in the transverse directions and $dE/dt$ is the average energy exchange rate between the thermostats and the thermostatic regions.

*AEMD simulations.* The AEMD technique [68, 69] is used to cross-validate the results from the NEMD method, and the ITC value reported in previous work [18]. After equilibrating the constructed Gr/$h$-BN bilayer model (see Figure S8), we switched to an NVE ensemble and applied two Langevin thermostats: one set to 350 K for Gr and another set to 250 K for $h$-BN. This step was performed for 1 ns to establish a stable temperature difference of $\Delta T_0 = 100$ K. After removing the thermostats, we monitored the temperature difference $\Delta T(t)$ over 2 ns, which decayed exponentially over time $t$ according to $\Delta T(t) = \Delta T_0 \exp(-t/\tau)$, where $\tau$ is the decay time to be determined (see Figure S8c). Based on the lumped heat-capacity model [68], the ITC can be extracted as

$$G = \frac{C_V}{S\tau}, \quad (3)$$

where $S$ is the cross-sectional area. Here, and $C_V$ is the effective constant-volume heat capacity of the Gr/$h$-BN bilayer system, calculated as $C_V = \frac{C_{\text{Gr}} \cdot C_{h\text{-BN}}}{C_{\text{Gr}} + C_{h\text{-BN}}}$, where both $C_{\text{Gr}}$ and $C_{h\text{-BN}}$ are given by $3Nk_{\text{B}}$, with $k_{\text{B}}$ being the Boltzmann constant.

In the NEMD and AEMD simulations, periodic boundary conditions were applied in the in-plane directions, while non-periodic boundary conditions were used along the heat transport direction. The MD simulations with the NEP model were performed using the GPUMD package [70] (version 3.9.2), while those using empirical potentials were carried out with LAMMPS [71] package (version 2 Aug 2023).

**Spectral Thermal Conductance and Quantum-Statistical Corrections.** Within the NEMD framework, one can first calculate the virial-velocity correlation function in the nonequilibrium steady state [72],

$$\boldsymbol{K}(t) = \sum_i \langle \mathbf{W}_i(0) \cdot \boldsymbol{v}_i(t) \rangle. \quad (4)$$

where $\boldsymbol{v}_i$ and $\mathbf{W}_i$ are the velocity and virial tensor of atom $i$, respectively. Then, the classical thermal conductance can be spectrally decomposed:

$$G(\omega, T) = \frac{2}{V\Delta T} \int_{-\infty}^{\infty} dt e^{i\omega t} K(t), \quad (5)$$

where $V$ is defined as the product of the cross-sectional area and the interlayer distance across the Gr/$h$-BN interface. With the classical spectral $G(\omega, T)$ available, a quantum-corrected spectral thermal conductance $G^q(\omega, T)$ can be obtained by multiplying it with a ratio between quantum and classical modal heat capacity [39–41],

$$G^q(\omega, T) = G(\omega, T) \frac{x^2 e^x}{(e^x - 1)^2}, \quad (6)$$

where $x = \hbar\omega/k_{\text{B}}T$, $\hbar$ is the reduced Planck constant, and $k_{\text{B}}$ is the Boltzmann constant. Then, the quantum-corrected thermal conductance is then obtained as an integral over the entire frequency range as

$$G^q(T) = \int_0^\infty \frac{d\omega}{2\pi} G^q(\omega, T). \quad (7)$$

For the quantum-corrected spectral thermal conductance of different stacking types in Gr/$h$-BN heterostructure, to prevent displacement of the stacking regions, a few atoms at the edges of the interfacial layers were fixed after optimizing the multilayer model. The regions corresponding to different stacking types in the two-layer heterointerface (see Fig. 3a and Figure S7) were then selected for calculation. In this case, the volume $V$ is calculated as the product of the area of the selected region and the average interlayer distance [31] between the two atomic layers. The reported spectral thermal conductance results for different stacking types (Fig. 5) are the averages of ten independent simulations.

**Phonon Spectral Energy Density Calculations.**
The phonon SED technique [73] is employed to examine the phonon dispersion and corresponding phonon lifetimes for the AA, AB, and ABp stacking configurations in the Gr/$h$-BN heterostructure. It represents the intensity of the kinetic energy distribution of a system over different phonon modes and can be calculated in MD simulations using the following formula [73]:

$$\Phi(\mathbf{q},\omega) = \frac{1}{4\pi\tau_0 N_T} \sum_\alpha^3 \sum_b^n m_b \left| \int_0^{\tau_0} \sum_l^{N_T} \dot{u}_\alpha(l,b,t) \right.$$
$$\left. \times \exp\left(i\mathbf{q}\cdot\mathbf{r}_0(l) - i\omega t\right) dt \right|^2. \qquad (8)$$

Here, $\tau_0$ is the total MD simulation time, while $n$ and $N_T$ represent the total basis atoms and the total unit cells in the system, respectively. The mass of the $b$-th basis atom is denoted by $m_b$, while $\dot{u}_\alpha(l,b,t)$ represents the $\alpha$-th Cartesian direction of the velocity of the $b$-th basis atom in the $l$-th unit cell at time $t$, and $\mathbf{r}_0(l)$ indicates the equilibrium position of the $l$-th unit cell.

The phonon lifetime time can be obtained by fitting the SED curve by the Lorentzian function [73],

$$\Phi(\mathbf{q},\omega) = \frac{I}{1 + [(\omega - \omega_c)/\gamma]^2}, \qquad (9)$$

where $I$ is the peak magnitude, $\omega_c$ is the frequency at the peak center, and $\gamma$ is the half-width at half-maximum. Finally, one can define the lifetime $\tau_{ph}$ at each wave vector $\mathbf{q}$ and frequency $\omega$ as:

$$\tau_{ph}(\mathbf{q},\omega) = \frac{1}{2\gamma}. \qquad (10)$$

Starting from the equilibrium configurations of the primitive cells for different stacking types (with interlayer distances of 3.56 Å for AA, 3.49 Å for AB, and 3.37 Å for ABp), a $6 \times 6 \times 50$ supercell was constructed for each configuration. All systems were first equilibrated at 300 K for 0.5 ns in the NVT ensemble, followed by a 3 ns production stage in the NVE ensemble. During this period, the velocities and positions of all atoms were collected every 50 fs, for use in the subsequent SED calculations. As the DW stacking represents a transitional state and cannot exist independently, it was excluded from the SED calculations. The SED calculation and Lorentzian fitting were performed using the PYSED package [74], while the lattice dynamics calculation was carried out with the PHONOPY package [75].

**Data availability:**

Complete input and output files for the NEP training are freely available at https://gitlab.com/brucefan1983/nep-data. Representative input and output files for different calculations in this work are freely available at https://github.com/Tingliangstu/Paper_Projects.

**Code availability:**

The source code and documentation for GPUMD are available at https://github.com/brucefan1983/GPUMD and https://gpumd.org, respectively. The source code and documentation for PYSED are available at https://github.com/Tingliangstu/pySED and https://pysed.readthedocs.io/en/latest/, respectively.

**Declaration of competing interest:**

The authors declare that they have no competing interests.

## CONTRIBUTIONS

TL proposed the initial idea, constructed the dataset, trained the NEP model, and performed all the MD calculations. TL and KX carried out the density functional theory calculations. TL and PY performed the benchmark tests. WJ calculated the van der Waals potential energy and interlayer distances. MH, XW, and ZY collected the experimental data. WO provided the initial training configurations. YY, XZ, ZY, ZF, and JX supervised the project. LT, KX, PY, ZF, and JX drafted the manuscript. All authors discussed the results and contributed to the writing of the manuscript.

## ACKNOWLEDGMENTS

T.L., K.X., and J.X. acknowledge the support from the National Key R&D Project from the Ministry of Science and Technology of China (Grant No. 2022YFA1203100), the Research Grants Council of Hong Kong (Grant No. AoE/P-701/20) and RGC GRF (No. 14220022). T.L. sincerely thanks for the Postgraduate Studentship from The Chinese University of Hong Kong. P.Y. is supported by the Israel Academy of Sciences and Humanities & the Council for Higher Education Excellence Fellowship Program for International Postdoctoral Researchers. W.O. acknowledges support from the National Natural Science Foundation of China (Nos. 12472099 and 12102307). Some of the computations were conducted at the Supercomputing Center of Wuhan University.

---

# Supplemental Material:

# Probing the ideal limit of interfacial thermal conductance in two-dimensional van der Waals heterostructures

Ting Liang[1], Ke Xu[1], Penghua Ying[2], Wenwu Jiang[3], Meng Han[4], Xin Wu[5], Wengen Ouyang[3], Yimin Yao[4], Xiaoliang Zeng[4], Zhenqiang Ye[*6], Zheyong Fan[†7], and Jianbin Xu[‡1]

[1]Department of Electronic Engineering and Materials Science and Technology Research Center, The Chinese University of Hong Kong, Shatin, N.T., Hong Kong SAR, 999077, P. R. China
[2]Department of Physical Chemistry, School of Chemistry, Tel Aviv University, Tel Aviv, 6997801, Israel
[3]Department of Engineering Mechanics, School of Civil Engineering, Wuhan University, Wuhan, Hubei 430072, China
[4]Shenzhen Institute of Advanced Electronic Materials, Shenzhen Institute of Advanced Technology, Chinese Academy of Sciences, Shenzhen 518055, PR China
[5]Institute of Industrial Science, The University of Tokyo, Tokyo 153-8505, Japan
[6]College of Materials Science and Engineering, Shenzhen University, Shenzhen 518055, China
[7]College of Physical Science and Technology, Bohai University, Jinzhou 121013, P. R. China

# Contents

[*]Email: yezq@szu.edu.cn
[†]Email: brucenju@gmail.com
[‡]Email: jbxu@ee.cuhk.edu.hk

# Supplemental Notes

## Supplemental Note S1: The INCAR input file for DFT-MD simulations

To obtain configurations with actual temperature fluctuations, we conducted DFT-driven MD simulations under an isothermal (NVT) ensemble at a target temperature linearly increasing from 10 to 1000 K within 20 ps. The DFT-MD simulations are performed with reduced computational precision, employing an energy cutoff of 500 eV and a convergence threshold of $10^{-4}$ eV. This relatively low precision setting, combined with Γ-point sampling in the Brillouin zone, is sufficiently accurate for structural sampling in generating training datasets while maintaining computational efficiency. We used the following inputs in the INCAR file of VASP (version 6.3.0) for DFT-driven MD simulations.

```
  GGA      = PE       # Use the PBE functional for exchange-correlation
  LREAL    = Auto     # Projection operators: Automatic
  ENCUT    = 500      # Cut-off energy for plane-wave basis set, in eV
  IVDW     = 202      # The many-body dispersion energy method
  IBRION   = 0        # Activate MD
  MDALGO   = 2        # 2=Nosé-Hoover, 3=Langevin dynamics algorithm
  ISIF     = 2        # 1=NVE, 2=NVT, 3=NPT
  ALGO     = VeryFast # Use RMM-DIIS algorithm for faster SCF convergence
  ISYM     = 0        # Disable symmetry, useful for AIMD simulations
  TEBEG    = 10       # Begin temperature (10 K)
  TEEND    = 1000     # Final temperature (1000 K)
  POTIM    = 2        # Timestep in fs
  NSW      = 10000    # Max ionic steps (20 ps in total)
  EDIFF    = 1E-04    # SCF energy convergence; in eV
  ISMEAR   = 0        # Gaussian smearing
  SIGMA    = 0.1      # Width of the smearing in eV
  PREC     = low      # Precision level: low
  NELM     = 200      # Max electronic SCF steps
  SMASS    = 1.0      # Fictitious mass (in amu) for lattice degrees of freedom
  NWRITE   = 1        # Control the verbosity of output
```

# Supplemental Note S2: The INCAR input file for reference data calculations

We employed the following inputs in the `INCAR` file of VASP (version 6.3.0) for reference data calculations.

```
GGA      = PE       # Use the PBE functional for exchange-correlation
LREAL    = Auto     # Projection operators: Automatic
ENCUT    = 850      # Cut-off energy for plane-wave basis set, in eV
IVDW     = 202      # The many-body dispersion energy method
KSPACING = 0.15     # Automatically calculate k-points
KGAMMA   = .TRUE.   # GAMMA point
NSW      = 1        # Default: NSW = 0
IBRION   = -1       # Ions are not moved
ALGO     = Normal   # This is the default
EDIFF    = 1E-08    # SCF energy convergence; in eV
ISMEAR   = -5       # Tetrahedron smearing method with Blöchl corrections
PREC     = Accurate # Precision level: high
NELM     = 150      # Max electronic SCF steps
```

# Supplemental Note S3: Hyperparameters for NEP training

We used GPUMD-v3.8 to train the NEP model, which is an NEP4 model [1]. The `nep.in` input file for GPUMD reads:

```
version     4              # NEP4
type        3   B  C  N    # Boron (B), Carbon (C), and Nitrogen (N)
cutoff      8 4            # Radial and angular cutoffs
n_max       8 8            # Size of radial and angular basis
basis_size  12 12          # Number of radial and angular basis functions
l_max       4 2 0          # Expansion order for angular terms
neuron      50             # Number of neurons in the hidden layer
lambda_e    1.0            # Weight of energy loss term
lambda_f    1.0            # Weight of force loss term
lambda_v    0.1            # Weight of virial loss term
population  50             # Population size used in SNES algorithm
batch       10000          # Batch size for training (Full batch)
generation  700000         # Number of generations used by SNES algorithm
```

The cutoff radii for radial and angular descriptor parts are set at $r_c^R = 8$ Å and $r_c^A = 4$ Å, respectively. To better capture the long-range dispersion interactions between the layers of the Gr/$h$-BN heterostructure, we employed a large cutoff radius for the radial descriptors. For the radial descriptor components, we utilized 9 radial functions, corresponding to `n_max + 1`, each represented as a linear combination of 13 basis functions (`basis_size + 1`). Similarly, the angular descriptor components also consist of 9 radial functions linearly combined from 13 basis functions (Chebyshev polynomials). For constructing the angular descriptor components, we incorporated three- and four-body correlations into the spherical harmonics, extending up to degree $l = 4$ and $l = 2$, respectively. Consequently, the total descriptor vector for a single element comprises $9 + 9 \times 5 = 54$ components. The neural network architecture for each element is structured as 54-50-1, with 50 neurons in a single hidden layer. For each pair of elements, there are $9 \times 13 + 9 \times 13 = 234$ trainable descriptor parameters, leading to a total of $(54 + 2) \times 50 \times 3 + 1 + 234 \times 3^2 = 10,507$ trainable parameters for our NEP model. The final NEP model was trained over 700,000 generations (steps) with a full batch size.

# Supplemental Note S4: Discussion of AEMD calculation results

In this work, the AEMD technique was also employed to calculate the ITC of the Gr/$h$-BN heterostructure. Compared to the NEMD method (Fig. S7), the AEMD technique demands only a bilayer model (Fig. S8) for ITC calculations in 2D materials and offers lower computational costs due to its transient nature [2, 3]. However, our calculations indicate that the AEMD method significantly underestimates the ITC of the Gr/$h$-BN heterostructure. Based on 10 independent NEP-AEMD simulations, the decay time $\tau$ is determined to be 113.1 ± 5.7 ps, and the ITC is deduced to be 6.9 ± 0.4 MW m$^{-2}$ K$^{-1}$. Employing AEMD driven by the Tersoff+LJ empirical potential, we report an ITC of 8.1 ± 0.9 MW m$^{-2}$ K$^{-1}$, which is very close to the result (approximately 5 MW m$^{-2}$ K$^{-1}$) calculated by Zhang et al. [4] using the same method and potential. Consequently, the ITC calculated by NEP-AEMD is significantly lower than that obtained by NEP-NEMD and experimental measurements (see Fig. S8d), possibly because some low-frequency phonons with long lifetimes (up to 1000 ps, as shown in Figure 6d of the main text) do not contribute to thermal conductance in the AEMD method [3]. Coincidentally, Sood et al. [5] used AEMD to calculate the ITC of the WSe$_2$/WS$_2$ heterostructure as 2.8 ± 1 MW m$^{-2}$ K$^{-1}$, which also demonstrates an underestimation relative to the experimental value (4.8 ± 1.6 MW m$^{-2}$ K$^{-1}$).

# Supplemental Table

## Supplemental Table S1: Benchmarking of lattice constant and interlayer spacing

Table S1: Comparison of equilibrium lattice constant and interlayer spacing of different structures calculated by the NEP model with those from DFT (PBE+MBD) and experimental measurements. Here, the lattice constant of monolayer graphene is denoted as $a_{\rm Gr}$, and the equilibrium interlayer spacing of graphite is denoted as $d_{\rm Graphite}$; the lattice constant of monolayer $h$-BN is labeled as $a_{h-\rm BN}$, and the equilibrium interlayer spacing of bulk $h$-BN is referred to as $d_{h-\rm BN}$; the lattice constant of bilayer Gr/$h$-BN heterostructure with ABp stacking (see Fig. 2a in the main text) is designated as $a_{\rm Gr/h-BN}$, and the interlayer spacing is labeled as $d_{\rm Gr/h-BN}$. Currently, there is no available experimental value for the lattice constant of the Gr/$h$-BN heterostructure.

| Method | DFT | NEP | Expt. |
|---|---|---|---|
| $a_{\rm Gr}$ (Å) | 2.46 | 2.46 | 2.46 [6, 7] |
| $d_{\rm Graphite}$ (Å) | 3.38 | 3.37 | 3.34 [8], 3.36 [9] |
| $a_{h-\rm BN}$ (Å) | 2.50 | 2.50 | 2.50 [10, 11] |
| $d_{h-\rm BN}$ (Å) | 3.31 | 3.26 | 3.33 [10–12], 3.30 [13] |
| $a_{\rm Gr/h-BN}$ (Å) | 2.48 | 2.48 | N/A |
| $d_{\rm Gr/h-BN}$ (Å) | 3.37 | 3.37 | 3.80 [13] |

# Supplemental Table S2: Summary of ITC data for different 2D heterostructures

Table S2: Summary of ITC data for various 2D heterostructures, including experimental measurements from Raman and TDTR techniques, as well as NEP-calculated ITC results for the Gr/$h$-BN system, as shown in Fig. 4a of the main text. Currently, no ITC results from MLP-driven NEMD calculations are available for other heterostructures. Despite the limited experimental ITC measurements, it is evident that the Gr/$h$-BN heterostructure displays the highest reported ITC, which aligns with the largest PDOS overlap illustrated in Fig. S12. For clarity, the PDOS overlap corresponding to different heterostructures is also presented in the table.

| 2D heterostructures | Simulations/Experiments | ITC (MW m$^{-2}$ K$^{-1}$) | PDOS overlap |
|---|---|---|---|
| Gr/$h$-BN | NEP (Classical) | 62.5 ± 10.8 | 8.46 |
|  | NEP (Quantum) | 58.4 ± 11.0 |  |
|  | Raman [14] | 52.2 ± 2.1 |  |
|  | TDTR [15] | $34.5^{+11.6}_{-7.4}$ |  |
|  | Raman [16] | 10 |  |
|  | Raman [17] | 7.4 ± 0.4 |  |
| Gr/MoS$_2$ | TDTR [18] | 27.0 | 2.23 |
|  | Raman [19] | 12 ± 3 |  |
|  | Raman [19] | 10 ± 3 |  |
| $h$-BN/MoS$_2$ | Raman [14] | 17 ± 0.4 | 2.06 |
| MoS$_2$/WSe$_2$ | Raman [19] | 9 ± 3 | 1.49 |
| $h$-BN/WSe$_2$ | TDTR [5] | 6.8 | 1.47 |
| $h$-BN/WS$_2$ | TDTR [5] | 6.8 | 1.32 |
| Gr/WSe$_2$ | Raman [19] | 5 ± 3 | 1.29 |

# Supplemental Figures

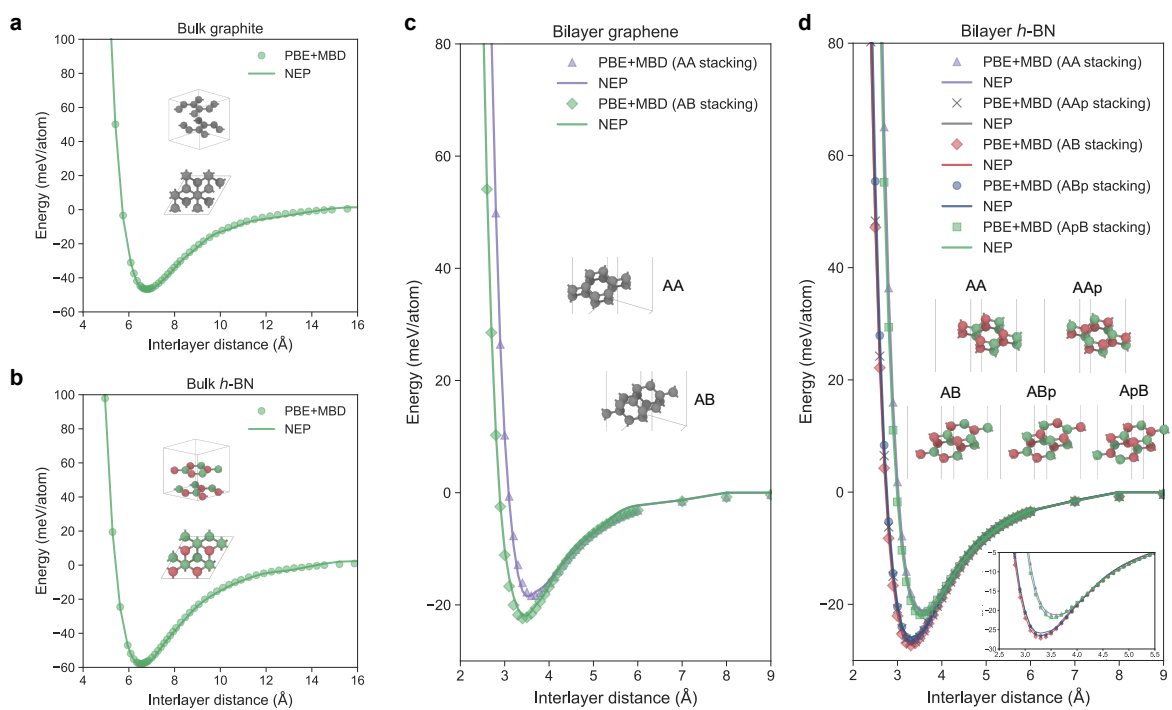

Figure S1: The binding energy curves corresponding to different stacking types for bulk and bilayer systems of Gr and $h$-BN, calculated using the NEP model and DFT (PBE+MBD). Snapshots of atomic models for various stacking types are shown.

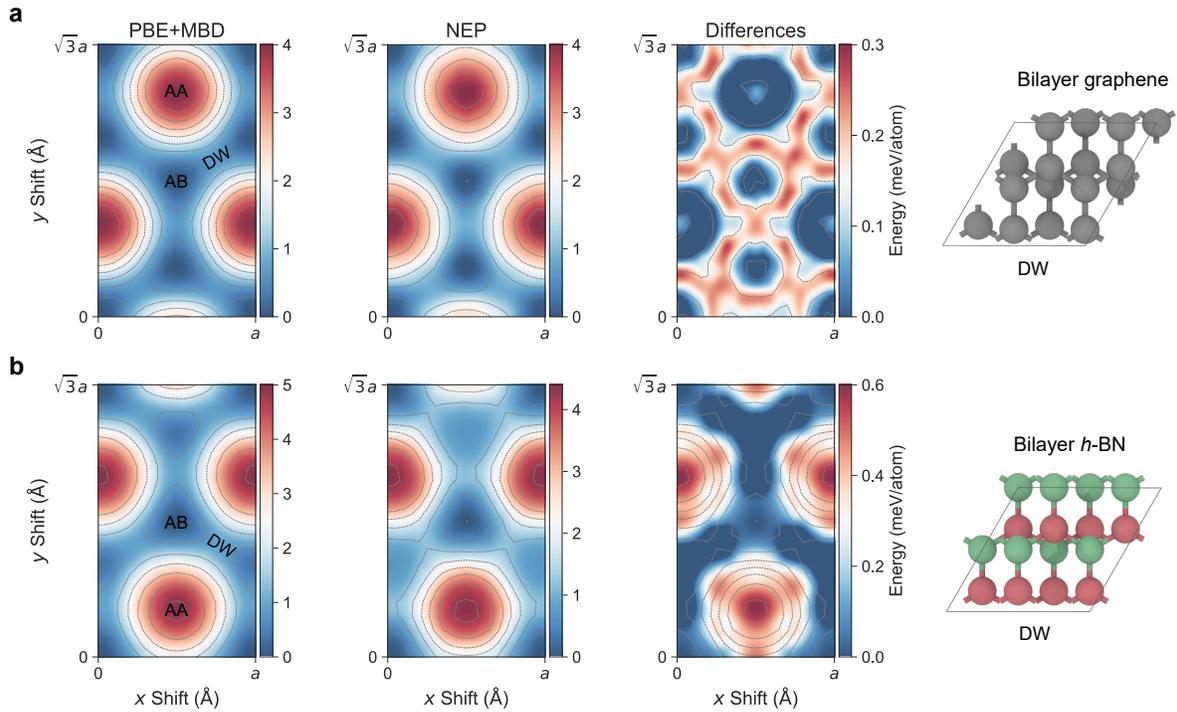

Figure S2: The sliding potential energy surfaces of **a** bilayer graphene and **b** bilayer $h$-BN predicted by the NEP model and DFT (PBE+MBD), and their differences. The AA, AB, and DW stacking modes are marked at the corresponding positions on the potential energy surfaces map. On the right, the atomic snapshots of the DW stacking type are visualized.

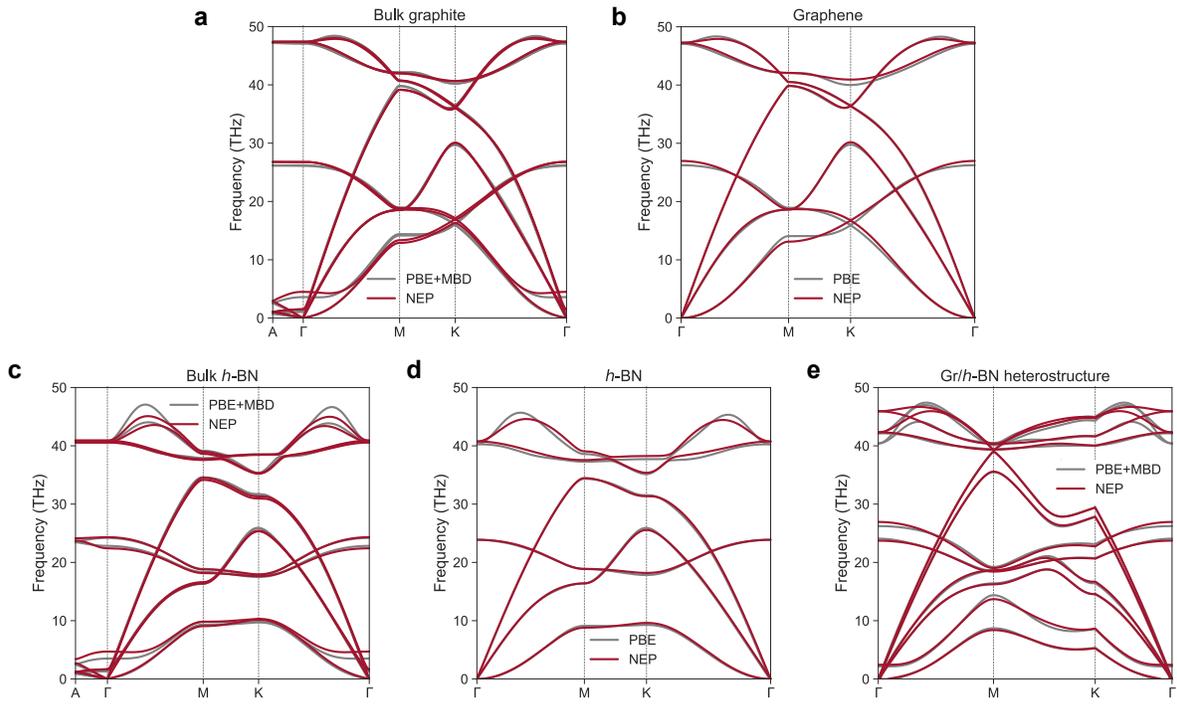

Figure S3: Comparison of the phonon dispersion predicted by DFT and NEP model for various systems.

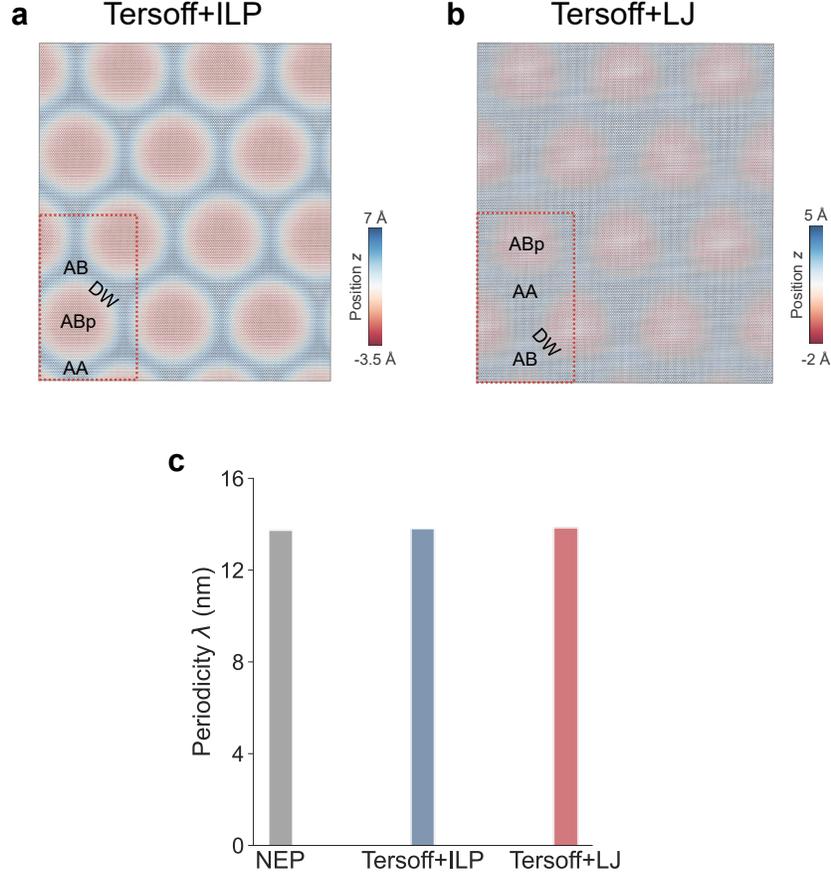

Figure S4: The moiré patterns and corresponding periodicities of the Gr/$h$-BN heterostructure generated by different potential models. **a** and **b** illustrate the moiré patterns generated by the combined empirical potentials, Tersoff+ILP [20–23] and Tersoff+LJ [20, 24], respectively. The regions corresponding to different stacking types are labeled for clarity. The red dashed box indicates the minimum periodic unit corresponding to the moiré pattern. **c** Comparison of periodicities of the Gr/$h$-BN heterostructure for different potentials. Although the periodicity of the moiré pattern from Tersoff+LJ is almost consistent with that of the NEP model and the Tersoff+ILP, the LJ term significantly underestimates the overall sliding energy landscape corrugation in Gr and $h$-BN [25], leading to substantial suppression of the height fluctuations of the moiré pattern.

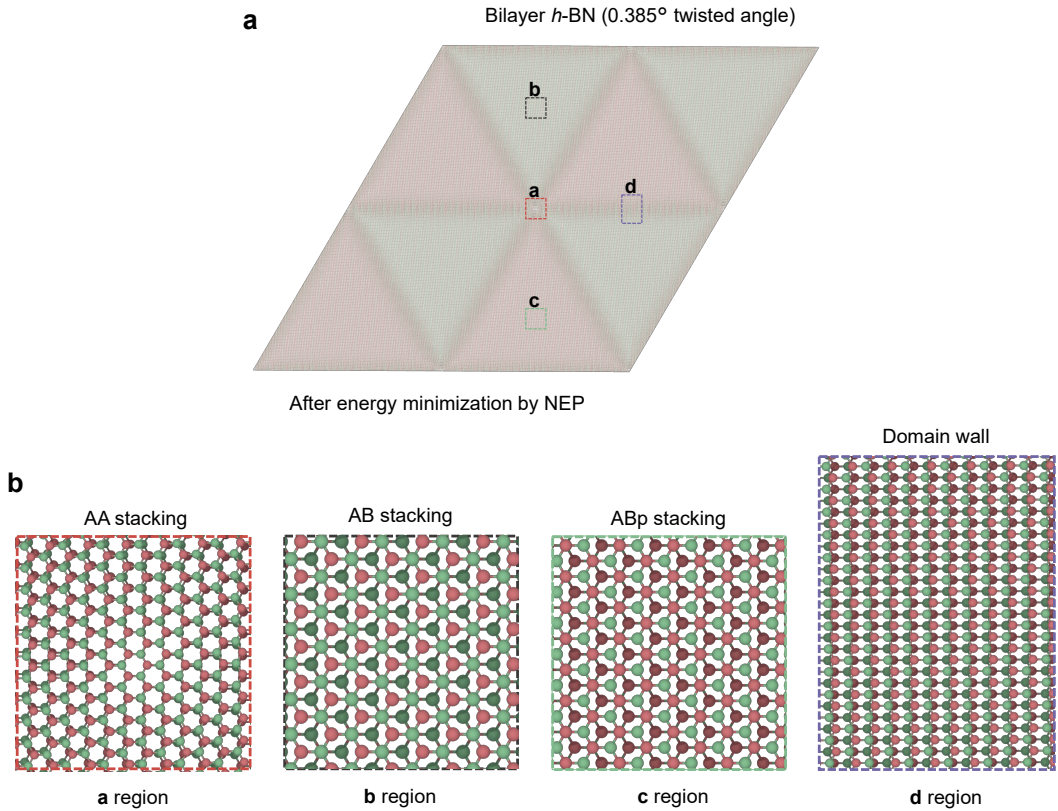

Figure S5: **a** Moiré pattern of twisted bilayer *h*-BN at an angle of 0.385° optimized by NEP model, and **b** its corresponding stacking types in different regions. The bilayer twist-angle model is based on the work by He et al. [26] on stacked engineered ferroelectrics, employing the deep potential (DP) model. Both the NEP and DP models, after the geometric optimization, accurately capture the experimentally observed moiré pattern in small-angle twisted bilayer *h*-BN [27, 28]. Despite the absence of twisted bilayer *h*-BN configurations in our training set, the NEP model successfully reproduced the twisted *h*-BN moiré pattern, showcasing its strong generalization ability.

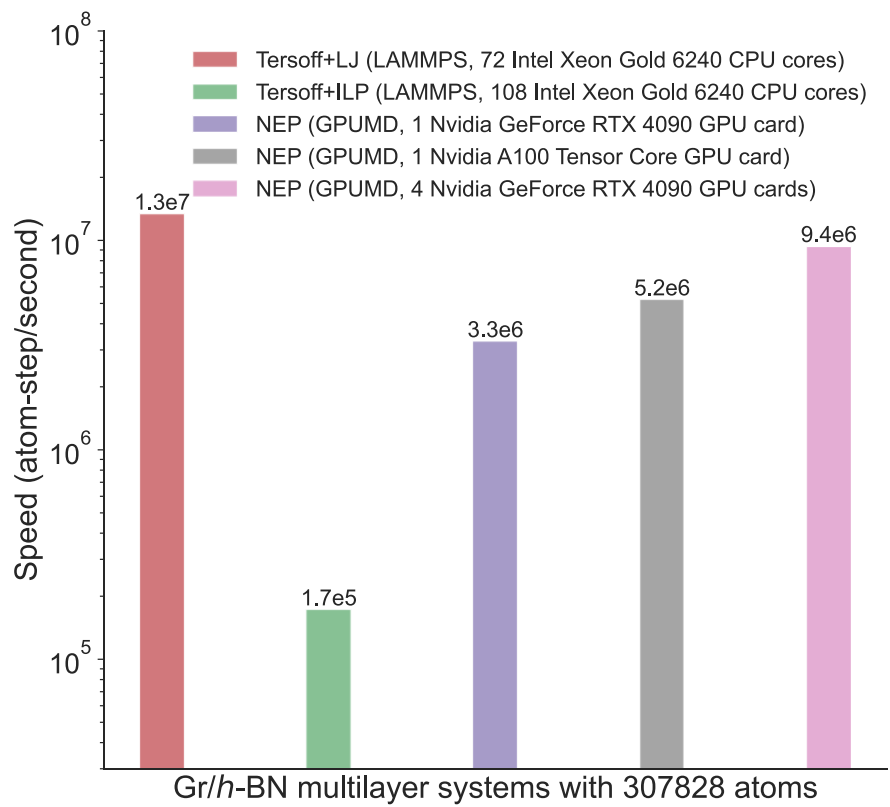

Figure S6: Comparison of the computational speeds of the NEP model and various empirical potentials. The computation speed is measured by running 1,000 steps of MD simulation in an isothermal ensemble, using a Gr/$h$-BN heterostructure system with 307,828 atoms (see Fig. S7a–b).

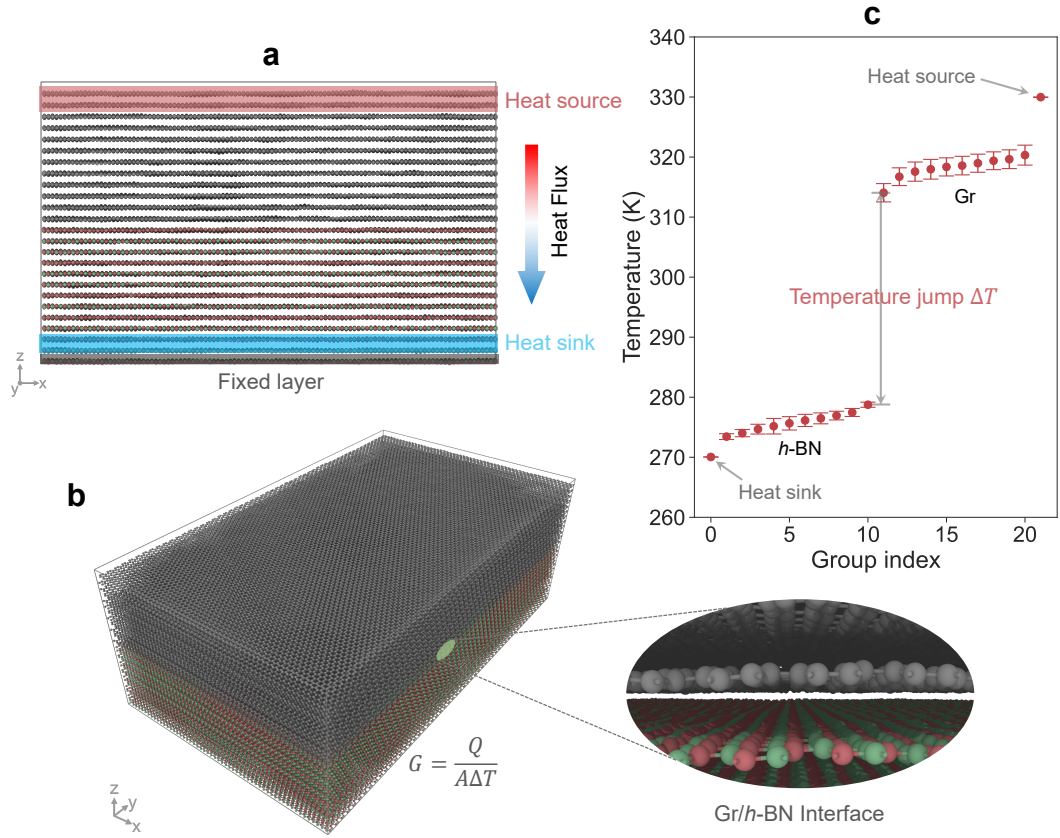

Figure S7: The schematic models for calculating the ITC of Gr/$h$-BN heterostructure using the NEMD method. **a** The side view of the atomic model for the multilayer Gr/$h$-BN heterostructure, consisting of a total of 25 layers. One layer of $h$-BN atoms is fixed, while the heat source and sink each occupy two layers to minimize excessive boundary temperature jump between the thermostatted regions and the rest of the system [29]. **b** The perspective view of the atomic model for the multilayer Gr/$h$-BN heterostructure. Here, $Q = dE/dt$ represents the energy transfer rate between the thermostat and the thermostated region. **c** A typical temperature profile of the Gr/$h$-BN heterostructure for NEMD calculations. Each individual layer is used to assess the temperature distribution statistically.

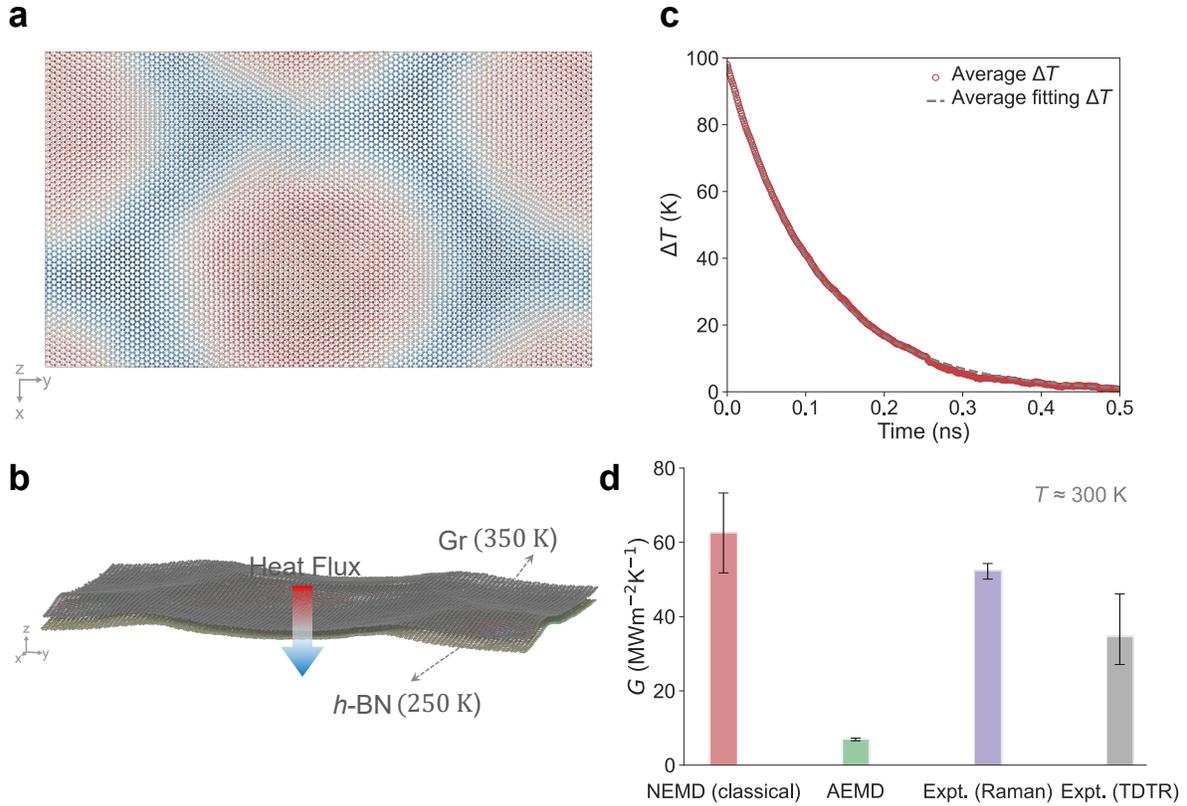

Figure S8: The schematic models for calculating the ITC of the Gr/$h$-BN heterostructure using the AEMD method. **a** A top view of the bilayer Gr/$h$-BN heterostructure atomic model after structural equilibrium at 300 K, showing distinct moiré patterns. The in-plane dimension of the bilayer model is consistent with that employed in the NEMD calculations. **b** The perspective view of the bilayer Gr/$h$-BN heterostructure atomic model along with the heat flux direction during the AEMD simulations. **c** The decay process of the average temperature difference between hot (Gr) and cold ($h$-BN) blocks. The gray dashed line indicates the exponential fitting, yielding the decay time $\tau$. Ten independent AEMD simulations were performed. **d** Comparison of the ITC of Gr/$h$-BN heterostructure calculated via the NEP model using AEMD and NEMD, alongside experimental measurements obtained from Raman [14] and TDTR [15] methods. The AEMD-derived ITC is notably lower than both the NEMD prediction and the experimental results.

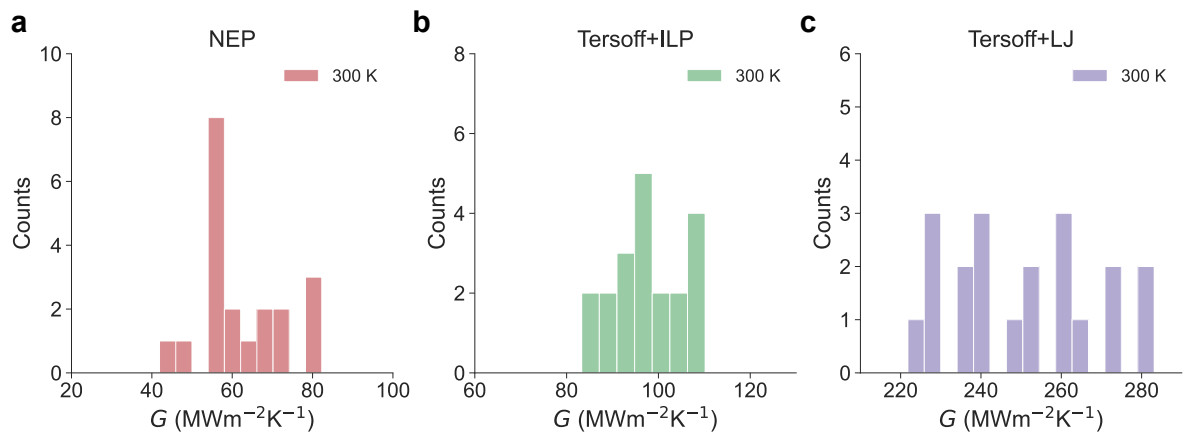

Figure S9: Distribution of ITC data from 20 independent calculations at 300 K for different potentials. The statistical error of ITC in the main text is the standard deviation of the average from the 20 independent runs.

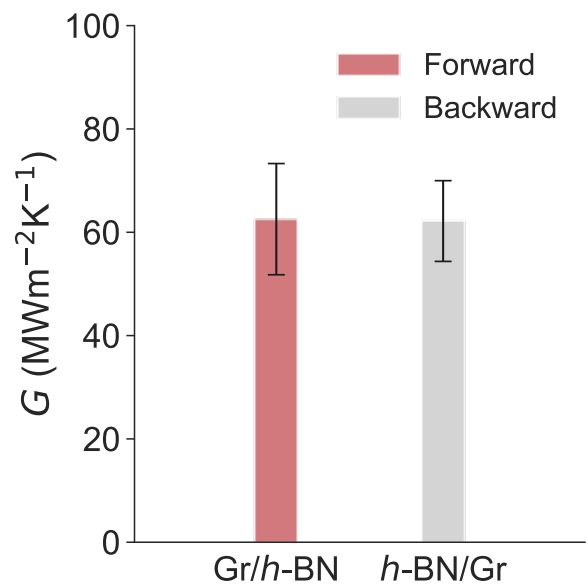

Figure S10: The ITC of the Gr/$h$-BN heterostructure calculated by swapping the heat source and sink in Fig. S7a. The ITC for forward and backward heat flow calculations is consistent within the error bars, indicating that the Gr/$h$-BN heterostructure system exhibits almost no thermal rectification, as demonstrated by Wu and Han [30].

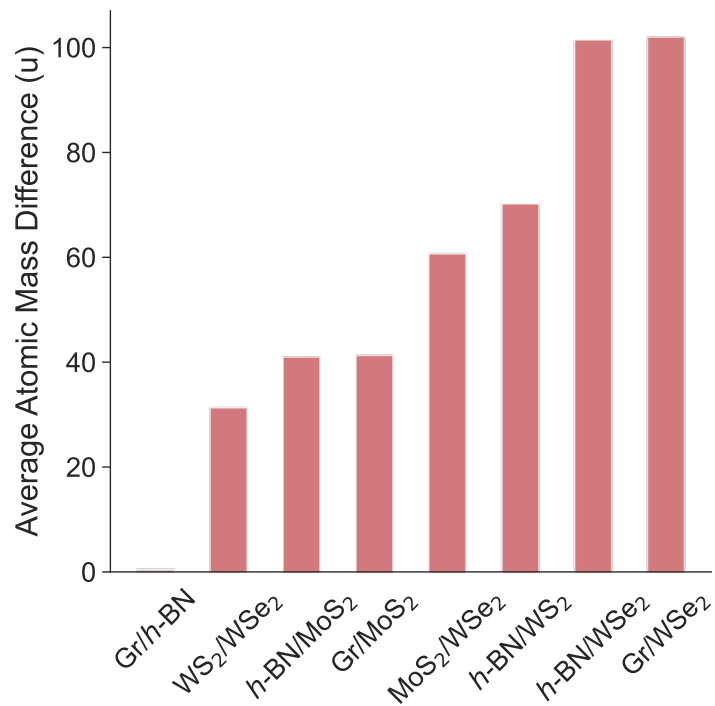

Figure S11: Average atomic mass differences of various heterostructures, in the unit of $u$. For each heterostructure, the atomic mass difference is calculated based on the average atomic mass per atom within the unit cell of each material. The Gr/WSe$_2$ heterostructure exhibits the greatest average atomic mass difference, underscoring the lowest energy transmission probability between them.

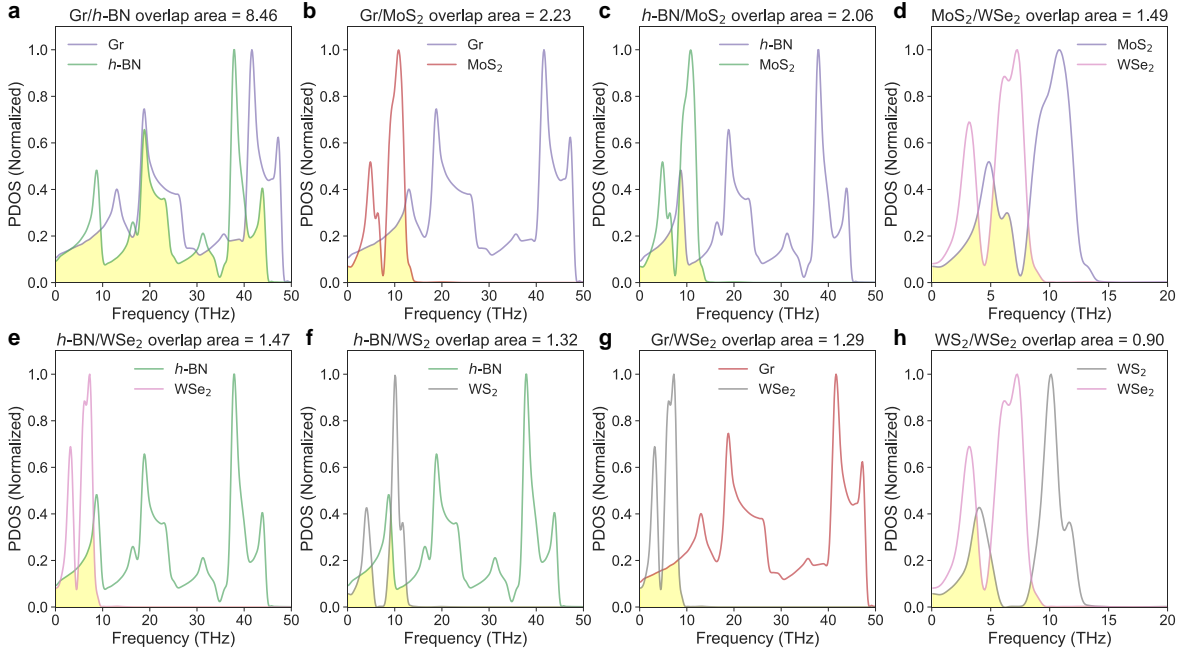

Figure S12: Normalized PDOS for different 2D heterostructure systems, with the overlapping regions highlighted in yellow. To quantitatively the PDOS mismatch of 2D heterostructures, the overlap area factor $S$ is defined as $S = \int_0^\infty \min\{P_A(\omega), P_B(\omega)\}\, d\omega$, where $P_A$ and $P_B$ are the two PDOS of different 2D materials. The calculation of PDOS and the corresponding overlapping area can also be found in Ref. [31]. Among different 2D materials, Gr and $h$-BN are the most similar in structure, mass, and thermal properties, and their overlapping area is the largest. For the $WS_2/WSe_2$ heterostructure with minimal PDOS overlap, despite not having the largest average atomic mass difference, it still corresponds to a low ITC value (about 4.8 MW m$^{-2}$ K$^{-1}$) [5]. This result is not presented in Fig. 4a of the main text, but it still supports the viewpoint that smaller PDOS overlap corresponds to lower ITC.

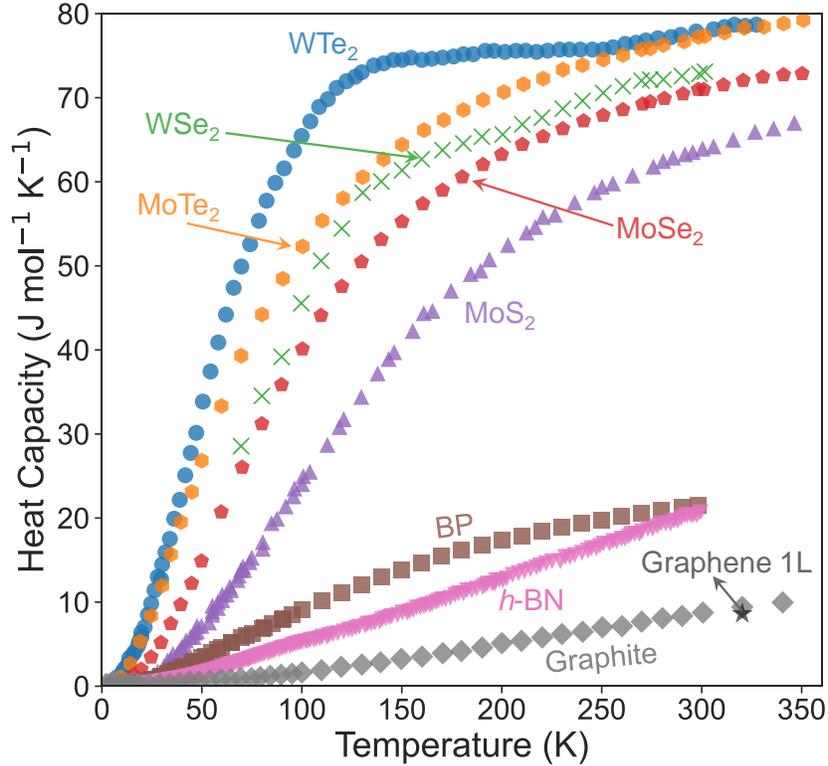

Figure S13: The molar heat capacity of bulk graphite (gray diamond) [32–34], h-BN (down-pink triangle) [35], black phosphorus (BP, brown square) [36, 37], MoS$_2$ (up-purple triangle) [38–40], MoSe$_2$ (red pentagon) [41], WSe$_2$ (green cross) [42], MoTe$_2$ (orange hexagon) [41], and WTe$_2$ (blue circle) [42], with the black star indicating single-layer graphene [43]. In crystalline solids, the classical heat capacity provides a reasonable approximation only at temperatures $T > \Theta_D$ (the Debye temperature), where all vibrational modes are fully excited. Below $\Theta_D$, specific heat strongly depends on temperature. As $\Theta_D$ is proportional to bond strength and inversely related to molecular mass, lighter layered materials like graphite and h-BN exhibit significantly higher $\Theta_D$ compared to transition metal dichalcogenides (TMDs). The TMDs tend to reach their Debye temperature, corresponding to their classical heat capacity limit of approximately 75 J mol$^{-1}$ K$^{-1}$, at a few hundred Kelvin, whereas graphite remains far below its classical limit, with a heat capacity of only about 8.3 J mol$^{-1}$ K$^{-1}$ at 300 K.

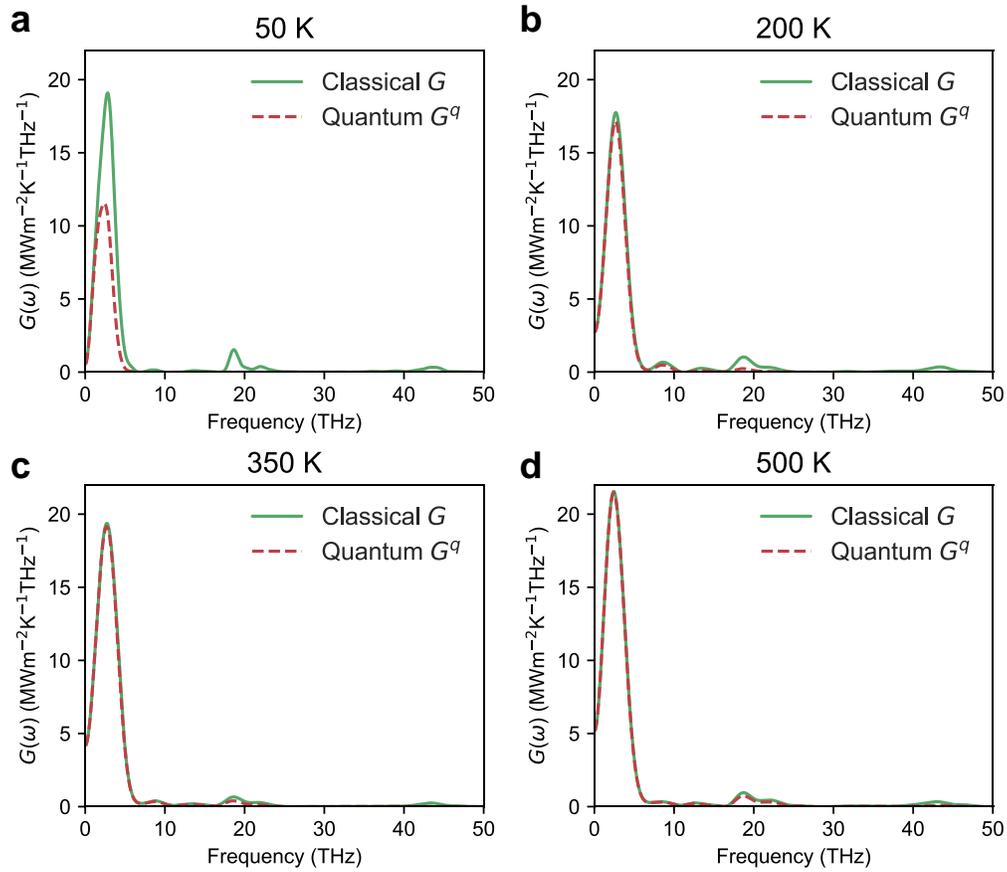

Figure S14: Classical and quantum-corrected spectral thermal conductance of Gr/$h$-BN heterostructure at temperatures of **a** 50, **b** 200, **c** 350, and **d** 500 K.

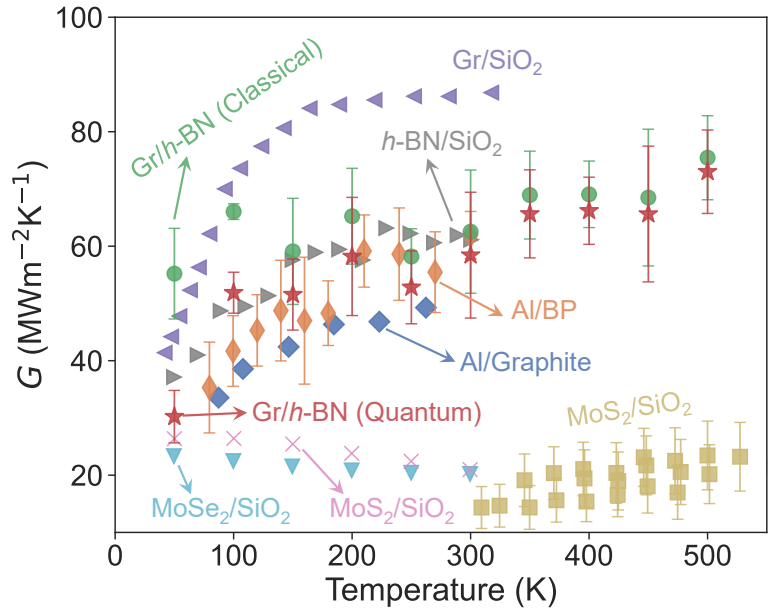

Figure S15: Temperature-dependent ITC ($G$) of different heterogeneous interfaces. The ITC results for the Gr/$h$-BN heterostructure are from the NEP model calculations, while those for other 2D/SiO$_2$ heterostructures, including Gr/SiO$_2$ [44], $h$-BN/SiO$_2$ [45], Al/BP [46], Al/graphite [47], MoS$_2$/SiO$_2$ [48, 49], and MoSe$_2$/SiO$_2$ [49], are based on experimental measurements.

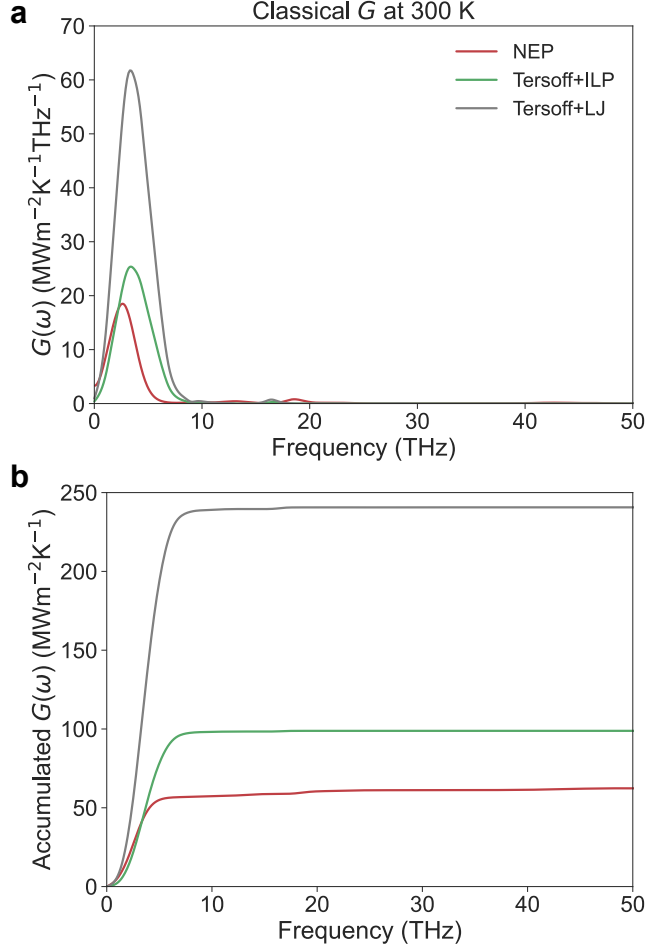

Figure S16: Comparison of **a** the classical spectral thermal conductance and **b** their corresponding accumulated thermal conductance calculated with different potential models. The classical spectral thermal conductance results for the empirical potentials Tersoff+ILP and Tersoff+LJ were calculated using Python code [50] combined with the LAMMPS [51] package, whereas the NEP results were derived using the GPUMD package [52]. The spectral thermal conductance results calculated using different potential models consistently indicate that the ITC of the Gr/$h$-BN heterostructure system is predominantly contributed by low-frequency phonons ($\omega/2\pi < 10$ THz), as long-wavelength, low-frequency phonons propagate more easily across the interface [53]. In the range below 10 THz, the accumulated spectral thermal conductance derived from the empirical potentials Tersoff+ILP and Tersoff+LJ is moderately and substantially higher, respectively, compared to the NEP-calculated results.

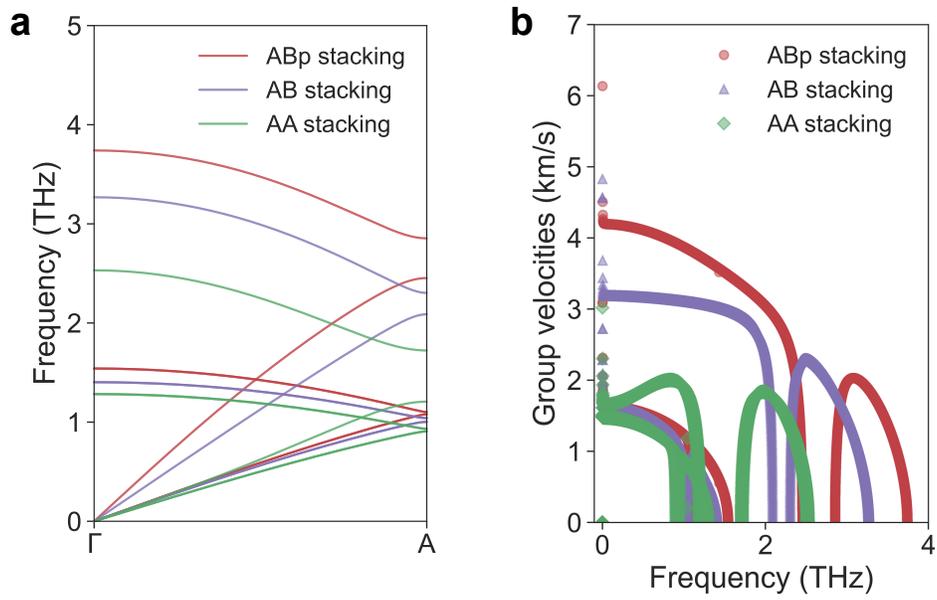

Figure S17: **a** and **b** represent the phonon dispersion (from $\Gamma$ to A) and the group velocity of the Gr/$h$-BN heterostructure for different stacking sequences.

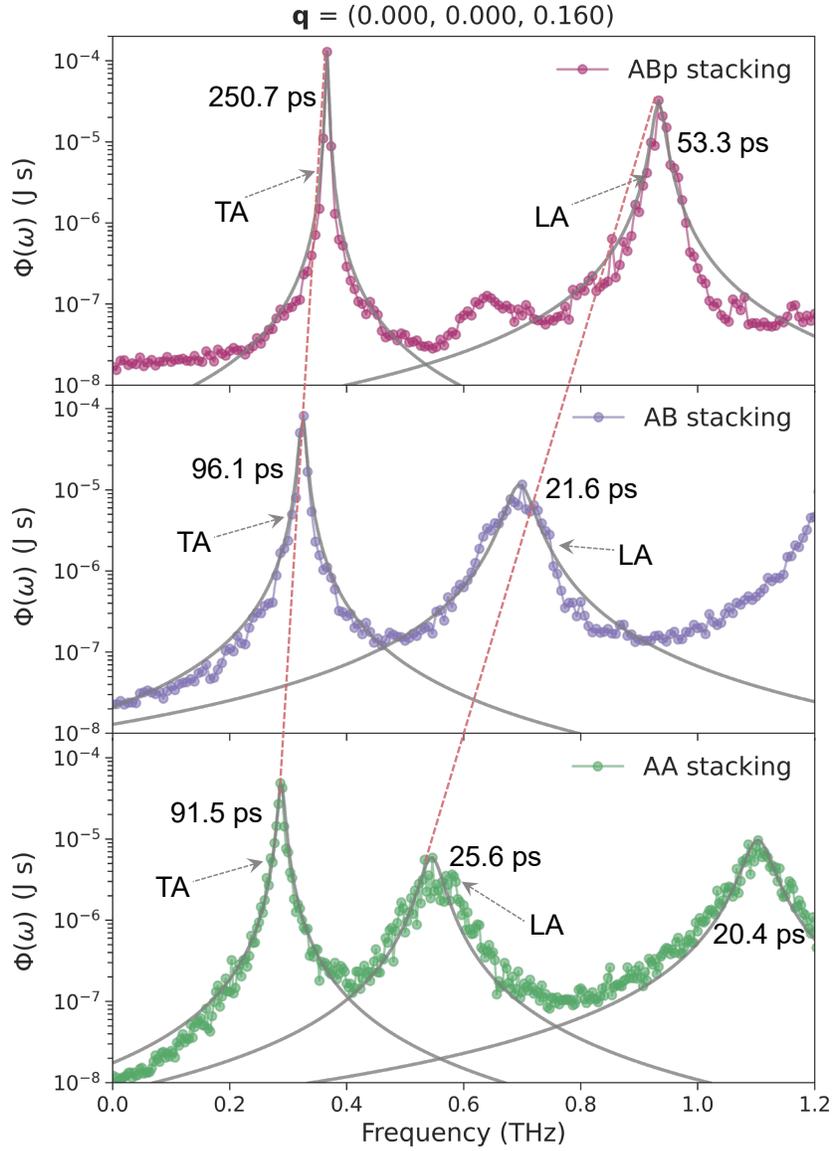

Figure S18: Lorentzian fitting of the SED peaks at a typical $q$ point. The phonon lifetimes obtained from fitting the TA and LA phonon branches for different stacking types are labeled. The grey solid line represents the quality of the Lorentzian fitting of the SED peaks, while the red dashed lines act as a visual guide to indicate the redshift of the TA and LA modes at the typical $q$ point transitioning from the ABp to the AA stacking type.

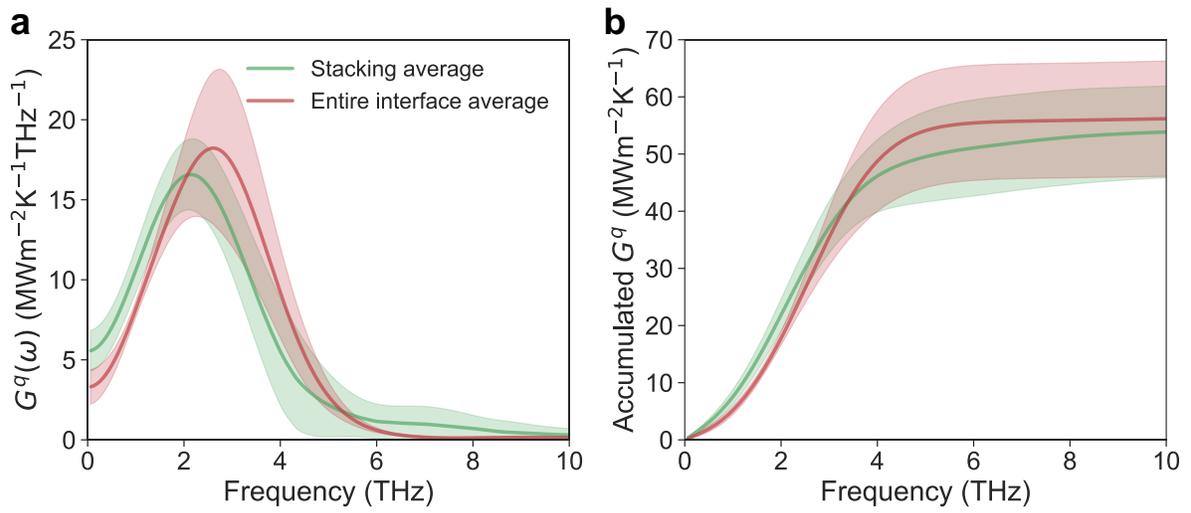

Figure S19: Comparison of **a** the quantum-corrected spectral thermal conductance and **b** the corresponding accumulated thermal conductance averaged over the entire heterointerface and that averaged over different stacking regions. The shading represents the standard deviation obtained from ten independent simulations. The accumulated quantum-corrected spectral thermal conductance obtained through two different averaging methods is nearly identical within the margin of error.

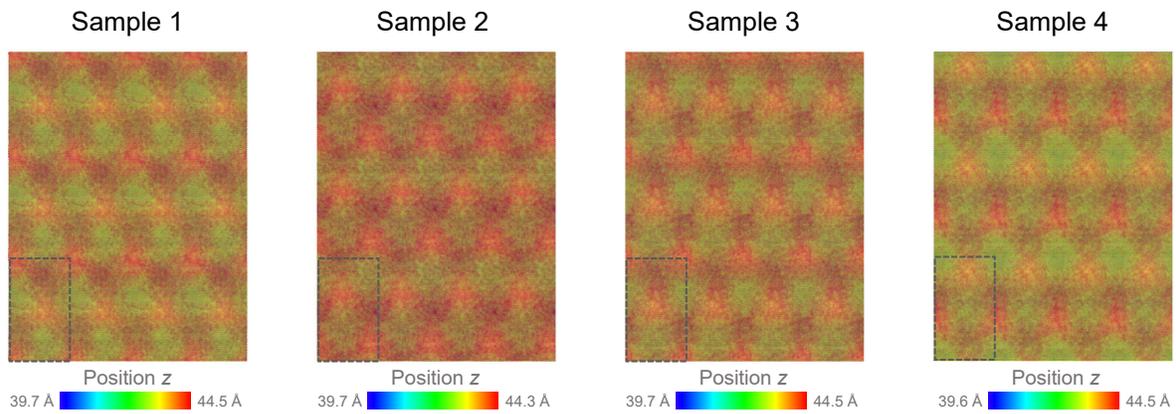

Figure S20: Atomic configuration snapshots at the interface of the multilayer Gr/$h$-BN heterostructure model (see Fig. S7) during NEMD simulations at 300 K. Four samples are taken from four independent simulations. The moiré patterns remain clearly visible at the heterointerface during the NEMD process. The black dashed box marks the minimal periodic unit corresponding to the moiré patterns.

# Supplemental References